\documentclass[aps,prc,nofootinbib,floatfix,superscriptaddress,longbibliography,reprint]{revtex4-2}

\usepackage{graphicx}
\usepackage[caption=false]{subfig}
\usepackage{dcolumn}
\usepackage{amsmath}
\usepackage{amssymb}
\usepackage{psfrag}
\usepackage{epsfig}
\usepackage{epsf}
\usepackage{float}
\usepackage{slashed}
\usepackage{wasysym}
\usepackage{xcolor}
\usepackage{multirow}
\usepackage{braket}

\usepackage{times}
\usepackage{txfonts}

\definecolor{linkcolor}{rgb}{0,0,0.40}
\usepackage[pdfpagelabels, pdfencoding=auto, psdextra]{hyperref}
\hypersetup{%
    pdfsubject = Paper,
    unicode = true,
    breaklinks = true,
    colorlinks = true,
    linkcolor = linkcolor,
    citecolor = linkcolor,
    menucolor = linkcolor,
    urlcolor = linkcolor
}

\usepackage[capitalize,sort]{cleveref}

\usepackage{siunitx}

\newcommand*{\elem}[2]{\ensuremath{{}^{#2}{\text{#1}}}}
\newcommand*{\NxLO}[1]{\ensuremath{
    \ifnum #1=0
        \text{LO}
    \else
        \ifnum #1=1
            \text{NLO}
        \else
            \text{N}^#1\text{LO}
        \fi
    \fi
}}

\newcommand*{\Nmax}{\ensuremath{N_\text{max}}}
\newcommand*{\Nmaxmax}{\ensuremath{\mathcal{N}_\text{max}}}

\begin{document}

\title{Precise neural network predictions of energies and radii from the no-core shell model}

\author{Tobias~Wolfgruber}
\email{tobias.wolfgruber@physik.tu-darmstadt.de}
\affiliation{Institut f\"ur Kernphysik, Fachbereich Physik, Technische Universit\"at Darmstadt, Schlossgartenstr. 2, 64289 Darmstadt, Germany}

\author{Marco~Kn\"oll}
\affiliation{Institut f\"ur Kernphysik, Fachbereich Physik, Technische Universit\"at Darmstadt, Schlossgartenstr. 2, 64289 Darmstadt, Germany}

\author{Robert~Roth}
\email{robert.roth@physik.tu-darmstadt.de}
\affiliation{Institut f\"ur Kernphysik, Fachbereich Physik, Technische Universit\"at Darmstadt, Schlossgartenstr. 2, 64289 Darmstadt, Germany}
\affiliation{Helmholtz Forschungsakademie Hessen f\"ur FAIR, GSI Helmholtzzentrum, 64289 Darmstadt, Germany}

\date{\today}

\begin{abstract}
For light nuclei, \textit{ab initio} many-body methods such as the no-core shell model are the tools of choice for predictive, high-precision nuclear structure calculations. The applicability and the level of precision of these methods, however, is limited by the model-space truncation that has to be employed to make such computations feasible.
We present a universal framework based on artificial neural networks to predict the value of observables for an infinite model-space size based on finite-size no-core shell model data. Expanding upon our previous ansatz of training the neural networks to recognize the observable-specific convergence pattern with data from few-body nuclei, we improve the results obtained for ground-state energies and show a way to handle excitation energies within this framework. Furthermore, we extend the framework to the prediction of converged root-mean-square radii, which are more difficult due to the much less constrained convergence behavior. For all observables robust and statistically significant uncertainties are extracted via the sampling over a large number of network realizations and evaluation data samples.
\end{abstract}

\maketitle

\section{Introduction}

The challenge of solving the nuclear many-body problem with realistic Hamiltonians is at the core of nuclear theory.
Throughout the past decades, the access to growing high performance computing (HPC) resources has revolutionized nuclear structure physics and caused a paradigm shift from computationally trivial to more demanding numerical models, giving rise to a range of successful \textit{ab initio} methods such as the no-core shell model (NCSM) \cite{BaNa13,NaQu09,Roth09,ZhBa93}, the coupled-cluster (CC) theory \cite{KoDe04}, self-consistent Green's function (SCGF) approaches \cite{DiBa04}, the in-medium similarity renormalization group (IM-SRG) \cite{HeBo16}, or quantum Monte Carlo (QMC) methods \cite{CaGa15}.
While they have proven to provide realistic results for a range of observables, their precision is limited by computational constraints.

In a concurrent development, modern machine learning algorithms have excelled in a variety of computational tasks such as pattern recognition, interpolation, or optimization and have naturally sparked the interest of researchers in many fields including nuclear structure physics.
This has led to various applications \cite{ClLi99,BoDi22} either through exploiting machine learning features as done in neural-network quantum state techniques \cite{KeRi20,AdCa21,GnAd22,LoAd22}, purely data driven approaches \cite{AtMa04,AkBa13}, or as supplemental tools for the aforementioned many-body methods \cite{NeVa19,JiHa19,KnoWo23}.
Focusing on the latter, machine learning is of particular interest when it comes to extending the reach of \textit{ab initio} methods by circumventing their computational limitations.
As most of them are based on a basis expansion of the many-body problem, a major restraint are the rapidly growing model spaces that quickly exceed the capabilities of even the largest HPC clusters available.
Due to the nature of \textit{ab initio} methods, which employ systematically improvable truncations, the errors induced by these truncations can, in principle, be controlled.
However, accurately quantifying them or extrapolating results from finite model spaces to the full many-body Hilbert space are extremely challenging tasks.
While ground-state energies are accessible with conventional extrapolation schemes based on exponential fits \cite{Roth09,MaVa09,BoFu08}, there are no widely established methods for other observables such as radii.

As shown in recent applications to NCSM or CC calculations of ground-state observables, artificial neural networks (ANNs) are a promising tool to tackle this challenge \cite{NeVa19,JiHa19,KnoWo23}.
The convergence behavior of these methods with respect to the model space size is controlled by a model-space truncation parameter and the harmonic-oscillator (HO) frequency of the underlying single-particle basis.

The ANN models for extrapolating such data separate into two conceptually different categories.
The first category attempts to emulate the functional dependence of the observable on the model-space size or, more precisely, the truncation parameter and the HO frequency, thus, directly replacing exponential or polynomial extrapolation methods with an ANN \cite{NeVa19,JiHa19}.
Approximations for the full Hilbert space are then given by the networks prediction for very large values of the model-space truncation parameter.
While this provides robust extrapolations for well balanced sets of training data, the applicability of such an ANN is, by construction, limited to the nucleus, interaction, state, and observable it was trained for. Thus, the computation of a large and balanced sets of training data as well as the ANN training process itself has to be repeated for each nucleus, interaction, state, and observable under consideration. This is a major bottleneck as the generation of training data becomes very costly. Moreover, the ANNs are used for a true extrapolation, i.e., an evaluation in an input parameter regime they were never trained for.

We have introduced a different approach in Ref.~\cite{KnoWo23}, which is more akin to a pattern recognition task.
In our ANN model, the observable in the full Hilbert space is directly predicted from a set of converging sequences in small model spaces, which share the same limit.
Hence, the ANN attempts to capture the convergence pattern and predict the converged value, similar to how an experienced practitioner would estimate it.
Since we need to specify the fully converged values for the training process, we can only use NCSM data from very light systems with $A\leq4$ for the training set. Despite this limitation, we can construct an extensive library of training data for a specific observable by varying the underlying Hamiltonian in addition to the nucleus and model-space parameters. In this way the training data encompasses very diverse convergence patterns so that, with a proper normalization, the ANN has learned all possible convergence patterns during the training. This makes this ANN universally applicable to different nuclei, interactions and states. This universality in a great advantage. The generation of training data and the training only happens once and the resulting ANNs are then simply applied with small sets of NCSM evaluation data for the systems of interest.

While our previous work \cite{KnoWo23} has demonstrated the capabilities of this universal ANN model for predicting ground-state energies, this paper focuses on improvements of the precision of the ANN prediction and the generalization to other observables, particularly excitation energies and root-mean-square (rms) radii. The description of these observables is more challenging because they show a much larger variety of non-monotonous convergence patterns, since they are not protected by the variational principle. We will introduce a simple normalization scheme that improves the precision of predictions of converged ground-state and excited-state energies and enables the robust prediction of converged radii. For all observables we will extract statistically meaningful uncertainties for the ANN predictions. We will demonstrate the application of the ANNs to a range of p-shell nuclei up to \elem{Be}{9}.

\section{No-Core Shell Model}

While these ANN approaches can be adapted to various many-body methods, we focus on the NCSM, which solves the many-body stationary Schrödinger equation as a matrix eigenvalue problem
\begin{align}
    \sum_j\braket{\phi_i|H|\phi_j}\braket{\phi_j|\psi_n} = E_n\braket{\phi_i|\psi_n} \quad \forall i,
\end{align}
by expanding the Hamiltonian $H$ as well as the eigenstates $\ket{\psi_n}$ in a complete basis. The basis $\{\ket{\phi_i}\}$ is a set of Slater determinants of single-particle basis states, in our case eigenstates of the HO. To numerically solve this matrix eigenvalue problem, a truncation to a finite model space is employed. The so-called $\Nmax$ truncation limits the total number of HO excitation quanta.
The convergence behavior as function of $\Nmax$ depends on the chosen HO frequency $\hbar\Omega$ of the single-particle basis, though in the limit $\Nmax \rightarrow \infty$ this dependency vanishes and the exact solution is recovered.

Observables other than the energies $E_n$ are easily accessible in the NCSM by constructing the matrix representation of the corresponding operator in the many-body basis and evaluating the expectation value with the calculated eigenvectors.
Since the NCSM is a variational method, the energy eigenvalues converge monotonously from above to the exact value with increasing $\Nmax$. This powerful constraint on the convergence pattern does, however, not hold for other observables, e.g.\ radii, leading to more complex and less constrained convergence patterns.

The Hamiltonian employed in the NCSM calculations consists of a kinetic energy contribution and inter-nucleon interactions, typically derived from chiral effective field theory (EFT). The are several families of chiral nucleon-nucleon (NN) plus three-nucleon (3N) interactions with different order in the chiral power-counting and different cutoffs available nowadays, which exhibit different NCSM convergence behaviors. Even for the softest, i.e., fastest converging, of the the interactions very large $\Nmax$ are required to reach convergence. Therefore, an additional pre-processing of the Hamiltonian through a similarity renormalization group (SRG) transformation is used to accelerate the model space convergence \cite{RoLa11,RoCa14,MaVa09,JuMa13,LENPIC21}. The SRG transformation introduces the continuous flow parameter $\alpha$ which provide direct control over the convergence rate.

For few-body systems, which include \elem{H}{2}, \elem{H}{3}, and \elem{He}{4}, we can use the Jacobi-NCSM \cite{NaKa00}, which is a efficient formulation of the NCSM for very light nuclei. It uses a relative basis instead of the typical $m$-scheme, which allows us to obtain fully converged calculations at minimal computational expense, ideal for generating large sets of training data. For heavier nuclei in the p-shell, as they are used as test cases for the network evaluation, we use the standard $m$-scheme NCSM.

\section{Artificial Neural Networks}

\begin{figure*}
    \centering%
    \subfloat[NCSM results]{
        \includegraphics[width=0.45\textwidth]{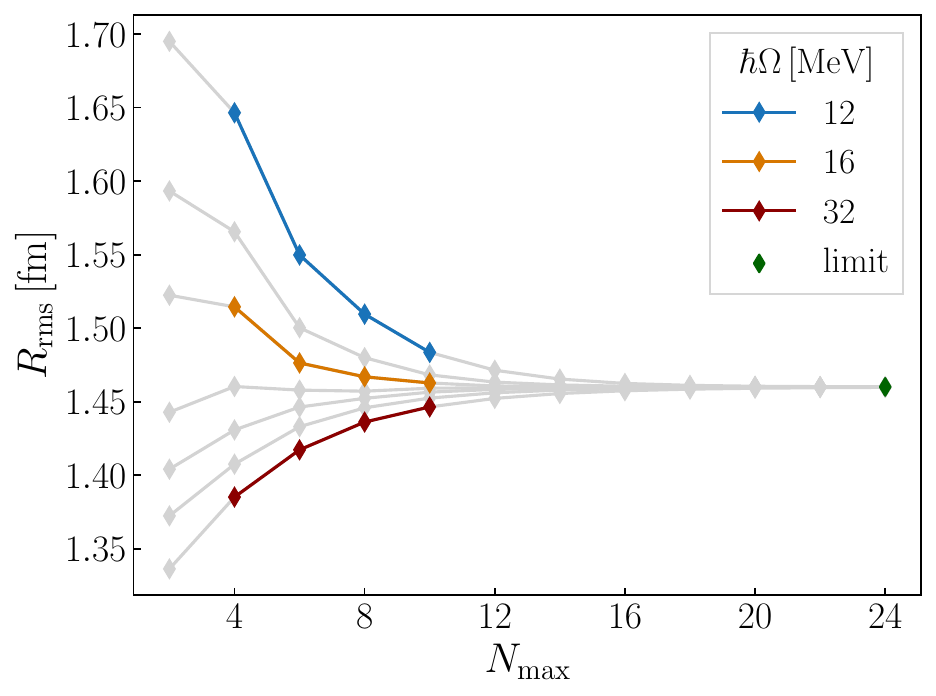}
        \label{fig:NCSM_ANN_subfigNCSM}}
    \qquad
    \subfloat[network topology]{
        \raisebox{0.2cm}{%
            \includegraphics[width=0.45\textwidth]{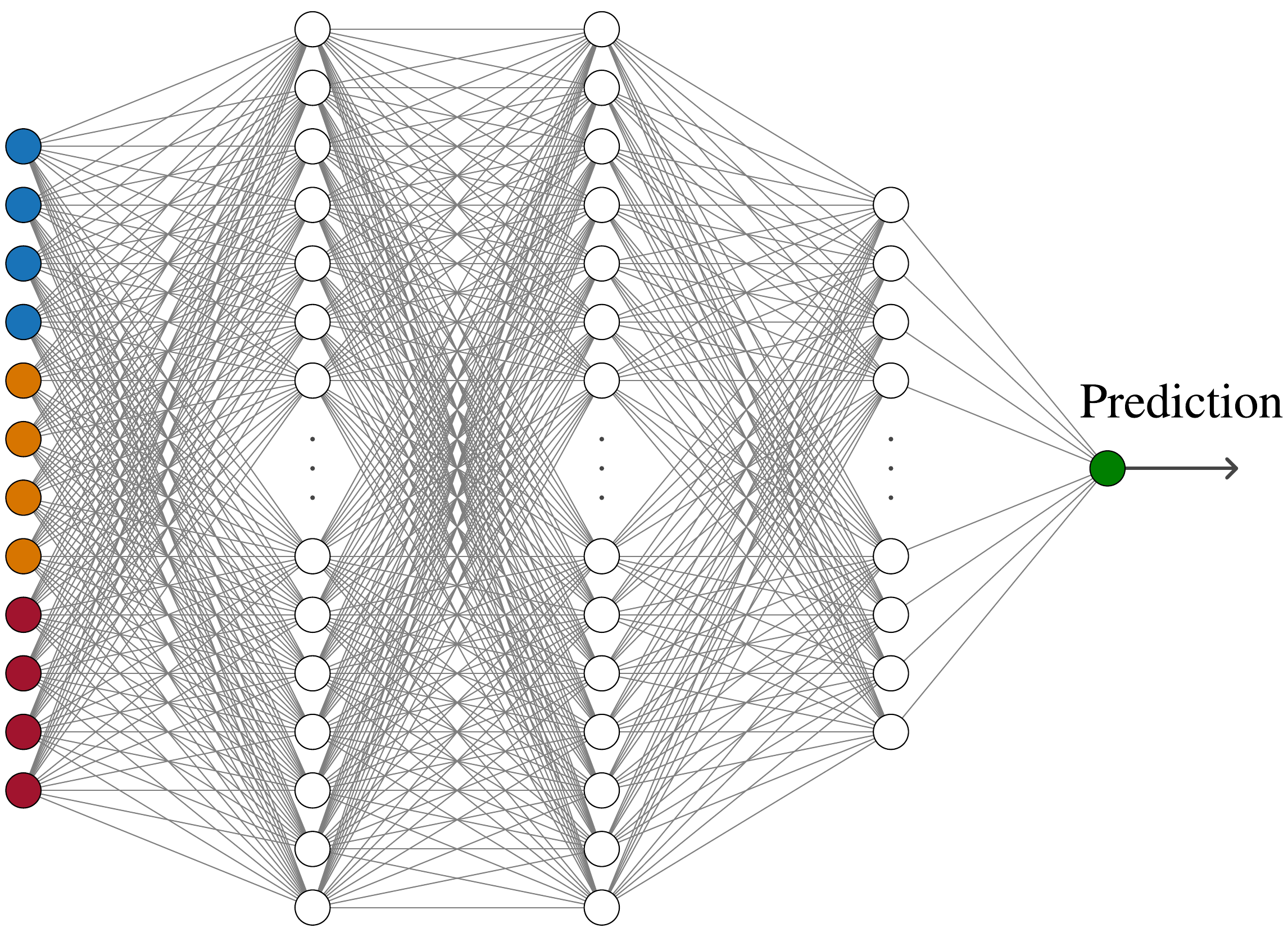}
        }
        \label{fig:NCSM_ANN_subfigANN}
    }
    \caption{(a) Exemplary Jacobi-NCSM results for the mass rms radius of \elem{He}{4}. All training data sequences for this combination of nucleus and interaction are shown in gray. Highlighted in color is one possible training sample, consisting of four radius results for consecutive $\Nmax$ values for three different HO frequencies, and the fully converged target value (green). (b) Schematic representation of the fully connected feed-forward network topology, where the colors of the input and output layer correspond to the colors of the sample in (a).
    }
    \label{fig:NCSM_ANN}
\end{figure*}

For our purpose, artificial neural networks are an ideal tool, as they excel at capturing patterns, especially in cases where classical algorithms are hard to implement or where the exact correlations between data points are unknown.
At their core, ANNs roughly emulate biological neural networks as present in the brain. The type of network we use is called a dense feedforward neural network. It consists of neurons with inputs and outputs, which are organized in layers. In this context dense means that each neuron's input is connected to the output of every neuron in the preceding layer. Based on these inputs, each neuron then calculates an output, which is fed to every neuron in the subsequent layer. For example, the $i$-th neuron in the $n$-th layer computes its output $x_n^{(i)}$ according to
\begin{align}
x_i^{(n)}= \sigma\left(\sum\limits_{j=0}^{N_{n-1}} x_{j}^{(n-1)} w_{ji}^{(n)} + b_j^{(n)}\right)\,,
\end{align}
where $N_{n-1}$ is the number of connected neurons from the previous layer, $w_{ji}^{(n)}$ and $b_j^{(n)}$ are the weights and biases which are optimized during training, and $\sigma$ is the activation function which rescales the output and in general introduces non-linearities.
The weights and biases are initialized randomly and are optimized during the training process via supervised learning. This means that every network input in the training data is labeled with the desired output.

The training process starts with a feed-forward pass, where a batch of training samples are evaluated and a combined deviation of the network's predictions from the desired results is calculated with a so-called loss function. The following back propagation step optimizes the weights and biases based on the gradient of the loss function. Repeating this process leads to an iterative improvement of the parameters of the network. A validation data set that is distinct from the training data is used to probe the performance of the network and the training progress and is usually used to adjust the step size in the gradient descent, commonly called learning rate. In then end, one should obtain a network that captures the patterns in the training data, leading to accurate predictions on the training data as well as previously unseen evaluation data.

So far we did not specify the choice of the so-called hyperparameters like the activation function, loss function, or back propagation algorithm. These choices can have a significant impact on the performance on both the training efficiency as well as on the quality of the networks themselves.
We achieved the best results with a rectified linear unit (ReLU) \cite{NaHi10} activation function, a mean-square error (MSE) loss function, and the AdamW back-propagation algorithm \cite{LoHu17} in combination with an adaptive learning rate scheduler, that halves the learning rate when the loss plateaus for two consecutive epochs, and a batch size of 512.
Note that, while the overall choice of hyperparameters is crucial for the performance of the ANNs, the predictions are very robust under reasonably small changes.

\section{Concepts and Network Design}

As already mentioned, artificial neural networks are ideally suited for situations where data are correlated, but where the patterns are either too varied or the problem is too hard to write down a functional dependence. This is exactly the case for NCSM calculations, where the convergence with $\Nmax$ clearly follows a certain pattern, that is hard to quantify.
The idea is to show the network multiple sequences of data points which all must converge to the same value and have the network output a prediction for the fully converged value of these sequences.
An illustration of converging NCSM sequences for the rms radius of \elem{He}{4} is shown in \cref{fig:NCSM_ANN_subfigNCSM}.

To construct such ANNs, we first need to train them on realistic NCSM convergence data. Since we use supervised learning, the exact, fully converged observable is a requirement for every sequence in the training data, which limits us to using few-body systems for the training as only these can be converged to the required precision.
However, since the convergence patterns in heavier p-shell nuclei are very similar to those in few-body systems, ANNs trained on the latter can provide precise predictions for a broad range of p-shell nuclei.
The main benefit of this method is that the training only needs to be done once and the trained ANNs can subsequently be applied, as a universal tool, to convergence sequences from any NCSM calculation.

Note that each observable needs its own set of networks, as the convergence pattern can vary significantly between different observables, though strongly related observables such as the ground-state energy and excited-state energies or rms radii and point-proton rms radii can use the same set of networks.

One important aspect we have not touched on yet is the exact design of the ANN including the structure of the data samples, which defines the input and output layers, and the topology, which determines the number of hidden layers.
For our application, the input for the network consists of $X$ sequences of NCSM calculations for different $\hbar\Omega$ which contain $L$ values of a specific observable for subsequent values of $\Nmax$.
This sets the input layer to be $X\cdot L$ neurons wide, while the output layer consists of a single neuron, which is supposed to be the prediction for the fully converged value of the observable.
We found that three hidden layers with sizes $4X\cdot L$, $4X\cdot L$, and $8X$ work very well for our use case. Fewer or much smaller hidden layers lead to less precise predictions, while adding more hidden layers or increasing their size yield diminishing returns, adding computational cost and increasing the risk of overfitting for only very marginal gains in precision. \Cref{fig:NCSM_ANN} visualizes both a training sample (a) and the network topology (b) for the input dimensions of $X=3$ and $L=4$, which is the structure used for all results presented in this work. Note that the input layer only accepts a subset of the available data in the NCSM sequences---there are typically more than $X=3$ different oscillator frequencies and more than $L=4$ different $\Nmax$ values available. This opens up the opportunity for a random sampling of the input data, which will be important for the training and statistical evaluation discussed below.

In order to evaluate the performance of our networks, it is helpful to compare them with traditional extrapolation techniques. For energy extrapolations we compare to an exponential fit scheme as described in Ref.~\cite{LENPIC21} for both the ground-state energy and excited-state energies. Radii are, due to being less constrained, more difficult to extrapolate. As there is no widely used algorithm for the extrapolation of radii, we forgo a comparison with a classical extrapolation technique for this observable.
An interesting point for a future comparison are the effective-theory-based infrared extrapolations for the NCSM \cite{WeFo15,FoCa18}. The applicability of these methods relies on specific and often untypical combinations of $\Nmax$ and $\hbar\Omega$, which require a large number of dedicated NCSM calculations and are not directly possible with the evaluation data used here.

\section{Training Data and Statistical Evaluation}

Our training data consists of $\elem{H}{2}$, $\elem{H}{3}$, and $\elem{He}{4}$ Jacobi-NCSM calculations up to $\Nmax=\numlist{50;40;24}$, respectively, for seven HO frequencies $\hbar\Omega=\qtylist{12;14;16;20;24;28;32}{MeV}$. We use the non-local NN+3N interactions from chiral EFT at orders $\NxLO{2}$, $\NxLO{3}$, and $\NxLO{4}'$ and for cutoffs $\Lambda=\qtylist{450;500;550}{MeV}$ introduced in \cite{EnMa17,HuVo20}. We applied these interactions bare and SRG evolved with three different flow parameters of $\alpha=\qtylist{0.02;0.04;0.08}{fm^4}$. All in all, we obtain 756 converging sequences. To reduce artifacts in the training set, we exclude $\Nmax=0$ data and remove sequences that are not sufficiently converged, which means that the value of the observable at the largest available $\Nmax$ shows a deviation of more than \qty{5}{\percent} from the fully converged value. The data that are left form the basis from which we can construct samples for training.
Each sample consists of a subset of the training data with a specific number of randomly selected HO frequencies and a specific window in  $\Nmax$, together with the fully converged value of the observable of choice.

We construct three disjoint subsets of samples: a large set of training samples, and two smaller sets of test and validation samples. The training consists of 20 epochs, iterating through all training samples with a batch size of 512 each time, monitoring the average loss via the test set at every cycle. This information is used to adaptively adjust the learning rate, which is initialized to 0.001. We evaluate the quality of the fully trained networks by measuring their performance on the validation set and discard poorly performing networks.

When applying the trained ANNs to unseen data we rely on a statistical evaluation for which we first construct evaluation samples analogously to the sample generation done for the training data, i.e., we construct all possible samples of $X=3$ HO frequencies and their permutations with values for the observable at $L=4$ consecutive $\Nmax$.
In addition to evaluating multiple samples, we also evaluate these samples with 1000 networks, that have been initialized differently and trained individually, in order to suppress any errors from a single ANN.
This multitude of predictions is then incorporated into a histogram that describes the distribution of the individual predictions, which usually shows a strong, single-peak structure that is very similar to a normal distribution.
For simplicity, we fit a Gaussian to the distribution to obtain the mean and standard deviation of the dominant peak structure.
While the extracted mean can be understood as the final ANN prediction for the given observable, the width of the distribution provides an uncertainty measure that incorporates not only all information from the evaluation data, but also accounts for the uncertainty of the individual ANNs.
Hence, we can provide precise predictions, which are robust against outliers, along with reliable $1\sigma$ uncertainties.

To be able to gauge the quality and consistency of the predictions, we group samples by the highest $\Nmax$ value they contain, denoted by $\Nmaxmax$.
Higher $\Nmaxmax$ correspond to larger model spaces and, therefore, better converged sequences, which should improve the ANN predictions of the fully converged value.
To emulate the situation in heavier p-shell nuclei where we see the largest potential for this method, we artificially limit the information for the evaluation in this work to $\Nmax=12$ or less. This corresponds to evaluation samples with $\Nmaxmax \le 12$.
With increasing $\Nmaxmax$ we expect the distribution of network predictions to decrease in width, corresponding to a smaller uncertainty, as well as improve in accuracy. Furthermore, the predictions should be compatible with each other within their uncertainty bands.

\section{Input Modes}

\begin{figure}[t]
    \centering
    \includegraphics[width=\columnwidth]{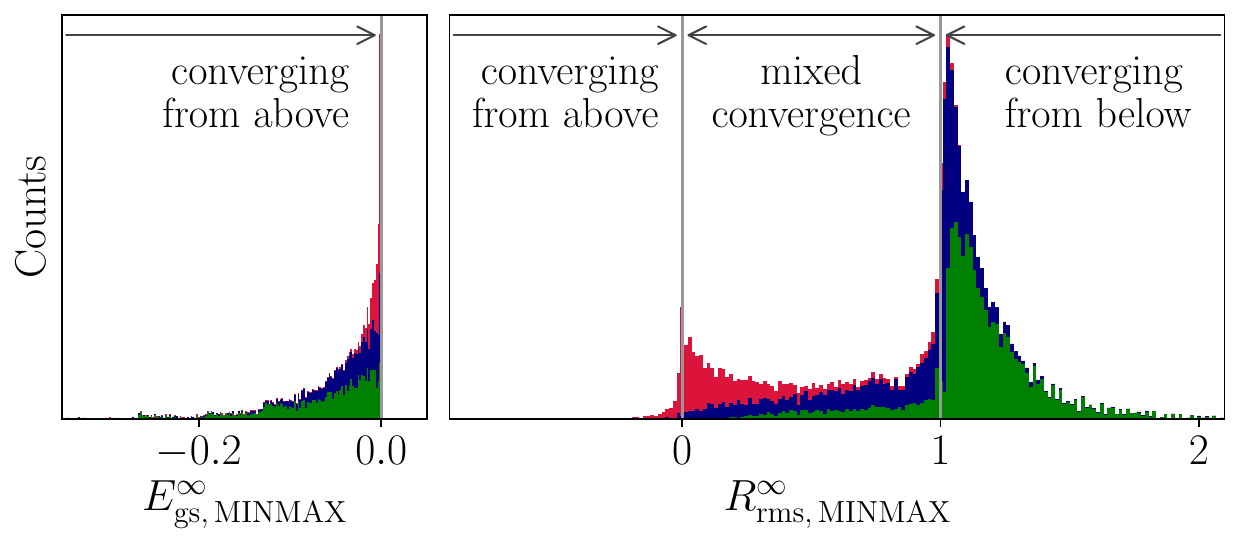}
    \caption{Histogram of the dimensionless target values of all samples in the MINMAX input mode for ground-state energies (left) and mass rms radii (right), separated by color: \elem{H}{2} (green), \elem{H}{3} (blue), and \elem{He}{4} (red). Values left of zero (right of one) belong to samples exclusively converging from above (below), while target values between zero and one belong to samples with a mixed convergence pattern.}
    \label{fig:MINMAX_target_distribution}
\end{figure}
\begin{figure*}[t]
    \includegraphics[width=0.9\textwidth]{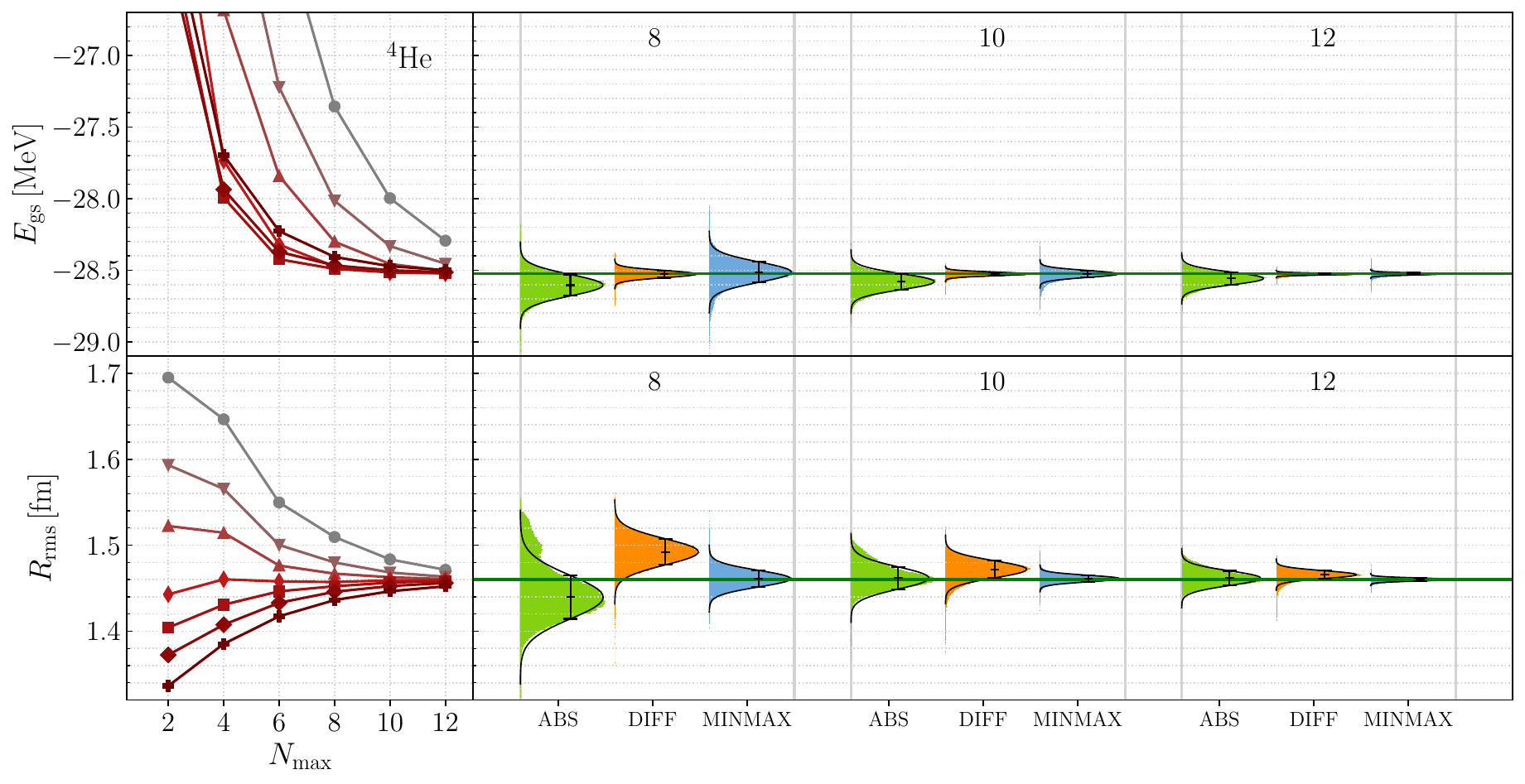}
    \caption{Input data and network predictions for the ground-state energy and mass radius of \elem{He}{4}. The left column shows the input data, consisting of NCSM calculations for $\hbar\Omega=\qtylist{12;14;16;20;24;28;32}{MeV}$ (gray to red). The right column shows a histogram of the predictions of the networks using the ABS, DIFF, and MINMAX input modes as as well as the fitted Gaussian and its mean and standard deviation. The three segments labeled \numlist{8;10;12} correspond to the $\Nmaxmax$ of the evaluation samples.}
    \label{fig:inputmodes}
\end{figure*}

There is some freedom when it comes to formatting the data that are fed into the input layer of the networks. We can formalize this by looking at the networks like a mapping $M:S\rightarrow O^\infty$, where $O^\infty$ is the converged observable and $S$ is the input sample. In our previous work \cite{KnoWo23}, we compared two distinct input modes. In the ABS mode, the data for a given observable, e.g.\ the ground-state energy, is simply fed in as-is. With our chosen network topology, a single ABS sample takes the shape
\begin{align}
\nonumber
S^{\Nmaxmax}_\text{ABS} = (&O_{\hbar\Omega_1}^{\Nmaxmax-6}, O_{\hbar\Omega_1}^{\Nmaxmax-4}, O_{\hbar\Omega_1}^{\Nmaxmax-2}, O_{\hbar\Omega_1}^{\Nmaxmax},\\
&O_{\hbar\Omega_2}^{\Nmaxmax-6}, O_{\hbar\Omega_2}^{\Nmaxmax-4}, O_{\hbar\Omega_2}^{\Nmaxmax-2}, O_{\hbar\Omega_2}^{\Nmaxmax},\\\nonumber
&O_{\hbar\Omega_3}^{\Nmaxmax-6}, O_{\hbar\Omega_3}^{\Nmaxmax-4}, O_{\hbar\Omega_3}^{\Nmaxmax-2}, O_{\hbar\Omega_3}^{\Nmaxmax})\,.
\end{align}
This is the naïve choice and gives the network all available information, but also the maximum amount of flexibility to look for patterns in the input data. A clear disadvantage is, however, that this can lead to undesired training results, as the networks just have to reproduce the few results for the nuclei in the training data without learning the convergence pattern.
Previous work has further shown, that this input mode induces a dependency on the respective energy range, which distorts the predictions in nuclei beyond the training set.
To circumvent this, we have introduced an additional step of scaling and shifting the data before feeding it to the network to increase variety in the target values and convergence patterns.
This slightly alleviates the effects of the large gap between the target value of the heaviest training nucleus and the heavier nuclei in the evaluation set.

The other mode we introduced in Ref.~\cite{KnoWo23} is called DIFF and feeds the differences between consecutive $N_\text{max}$ steps into the networks. This inherently alleviates the difference in scales between the nuclei in the training and the evaluation set, though shifting the data obviously had no more effect so that one source for inflating the initial training data pool was eliminated.
Both of these modes were explored with the ground-state energy as the target observable (for details see Ref.~\cite{KnoWo23}). Unfortunately, both exhibit significant shortcomings when applied to an observable with a different convergence structure like the radius.

This leads us to look into another input mode called min-max normalization, which we refer to as MINMAX. The min-max normalization is a common normalization technique for machine learning applications. In general, it normalizes the input data per sample to an interval $[a,b]$, though we exclusively choose an interval of $[0,1]$ for our network input. A single MINMAX sample can be defined via the ABS sample and takes the form
\begin{align}
S^{\Nmaxmax}_\text{MINMAX} = \frac{S^{\Nmaxmax}_\text{ABS} - \min(S^{\Nmaxmax}_\text{ABS})}{\max(S^{\Nmaxmax}_\text{ABS}) - \min(S^{\Nmaxmax}_\text{ABS})}\,.
\end{align}
To recover the value of the predicted observable $O^\infty$ from the network output $O_\text{MINMAX}^\infty$, the inverse transformation has to be applied:
\begin{align}
\hspace{-3pt} O^\infty = O^\infty_\text{MINMAX} \left( \max(S^{\Nmaxmax}_\text{ABS}) - \min(S^{\Nmaxmax}_\text{ABS}) \right) + \min(S^{\Nmaxmax}_\text{ABS}) .
\end{align}
The goal is to get rid of scale dependencies and reduce the information to the network in such a way that the recognition and extrapolation of the convergence pattern is the only viable strategy for the network due to a lack of other clues or reference points contained within the data.

\Cref{fig:MINMAX_target_distribution} shows the distribution of the MINMAX-normalized target values of all training samples for ground-state energies and radii. The colors correspond to the three training nuclei.
Both normalized target values are dimensionless, and the position on the $x$-axis determines if the value belongs to a sample that converges exclusively from above (below 0), exclusively from below (above 1), or one that has a mixed convergence (between 0 and 1).
As ground-state energies are constrained by the variational principle, all samples must naturally converge from above.
Furthermore, the ground-state energy usually converges rather fast, leading to most target values being grouped closely below zero.
Radii on the other hand are, as previously mentioned, not very well constrained.
The direction of convergence mainly depends on the specific HO frequency in conjunction with the nucleus and the $\Nmax$ window.
For both \elem{H}{2} and \elem{H}{3}, samples that converge from below dominate in our training set, while \elem{He}{4} appears to be much more balanced in that regard.
However, we expect this to be a good match for the situation of most nuclei we are interested in, as a typical energy-guided selection of HO frequencies leads to the majority of sequences for the rms-radius to converge from below.
For the training, we generate \num{1000000} training samples for the ABS and DIFF input modes via random scaling and shifting, while we are limited to about \num{350000} native samples for MINMAX.

The performance of the MINMAX input mode is compared to ABS and DIFF in \cref{fig:inputmodes} for the \elem{He}{4} ground-state energy and mass rms-radius. The left-hand panels show the evaluation data for both observables, while the right-hand panels shows histograms of the networks' predictions together with the fitted Gaussian for all three input modes at increasing $\Nmaxmax$. The fully converged value obtained from large $\Nmax$ is depicted as a horizontal green line.

Starting with the ground-state energies, all three input modes perform well, though ABS gives slightly overbound predictions, while both DIFF and MINMAX are very accurate even at $\Nmaxmax=8$. DIFF gives very small uncertainties compared to both other modes, though the MINMAX uncertainties decrease significantly with increasing $\Nmaxmax$ down to approximately the level of DIFF, while the ABS uncertainties remain rather large. As the DIFF uncertainties can tend to be somewhat underestimated, MINMAX seems to be a very good alternative, retaining the accuracy of DIFF with the slightly more realistic uncertainties.
The radius results in the lower panel show a much larger discrepancy between the different modes. Both ABS and DIFF give rather inaccurate results, especially for smaller $\Nmaxmax$. The uncertainties of both modes also remain fairly large. The MINMAX input mode on the other hand is already extremely accurate at $\Nmaxmax=8$, and subsequent $\Nmaxmax$ increase the precision in a realistic and systematic way when compared to the raw data on the left.

Though not shown here, the different input modes show a very similar behavior for other nuclei as well. This means that MINMAX is clearly superior for radii, and at least comparable if not better than both other input modes for ground-state energies. Therefore, we will exclusively use the MINMAX input mode for all following investigations.

\section{Results for Ground-State and Excitation Energies}
\begin{figure}[t]
    \includegraphics[width=1\columnwidth]{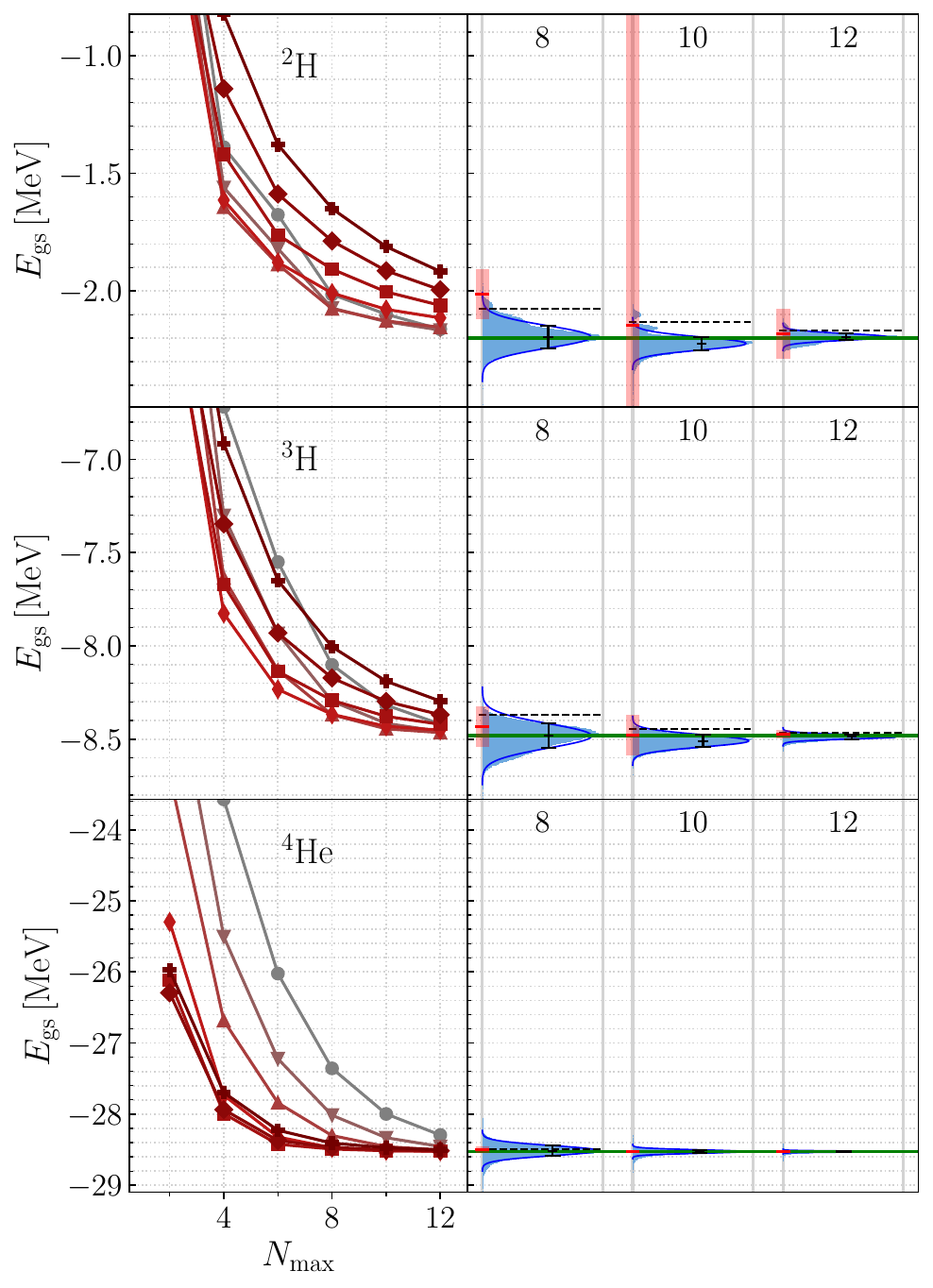}
    \caption{Input data and network predictions of the ground-state energy for \elem{H}{2}, \elem{H}{3}, and \elem{He}{4}. The left hand panels show the NCSM input data for $\hbar\Omega=\qtylist{12;14;16;20;24;28;32}{MeV}$ (gray to red), while the right hand panels show a histogram of the network predictions together with a fitted Gaussian, which gives the mean and 1-$\sigma$ uncertainty. The variational minimum at each $\Nmaxmax$ is given by the black dashed lines, the classical extrapolation results in red with the associated uncertainty, and the fully converges value is given by the green line.}
    \label{fig:LightEgs}
\end{figure}
\begin{figure}[t]
    \includegraphics[width=1\columnwidth]{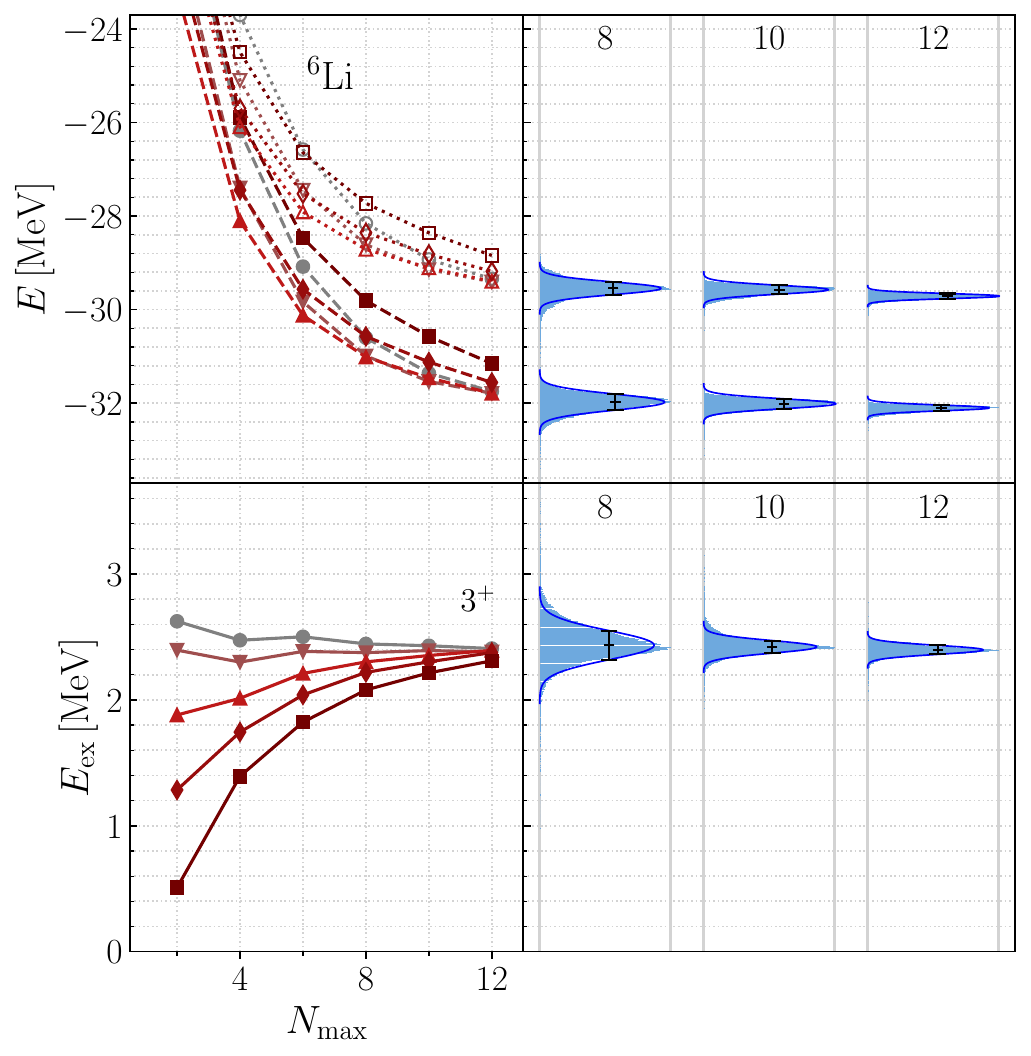}
    \caption{Input data and network predictions for the $1^+$ ground and $3^+$ excited state of \elem{Li}{6} (upper panel) and for the excitation energy of the $3^+$ excited state (lower panel) with $\hbar\Omega=\qtylist{14;16;20;24;28}{MeV}$ (gray to red).}
    \label{fig:E*Example}
\end{figure}
\begin{figure}[t]
    \includegraphics[width=1\columnwidth]{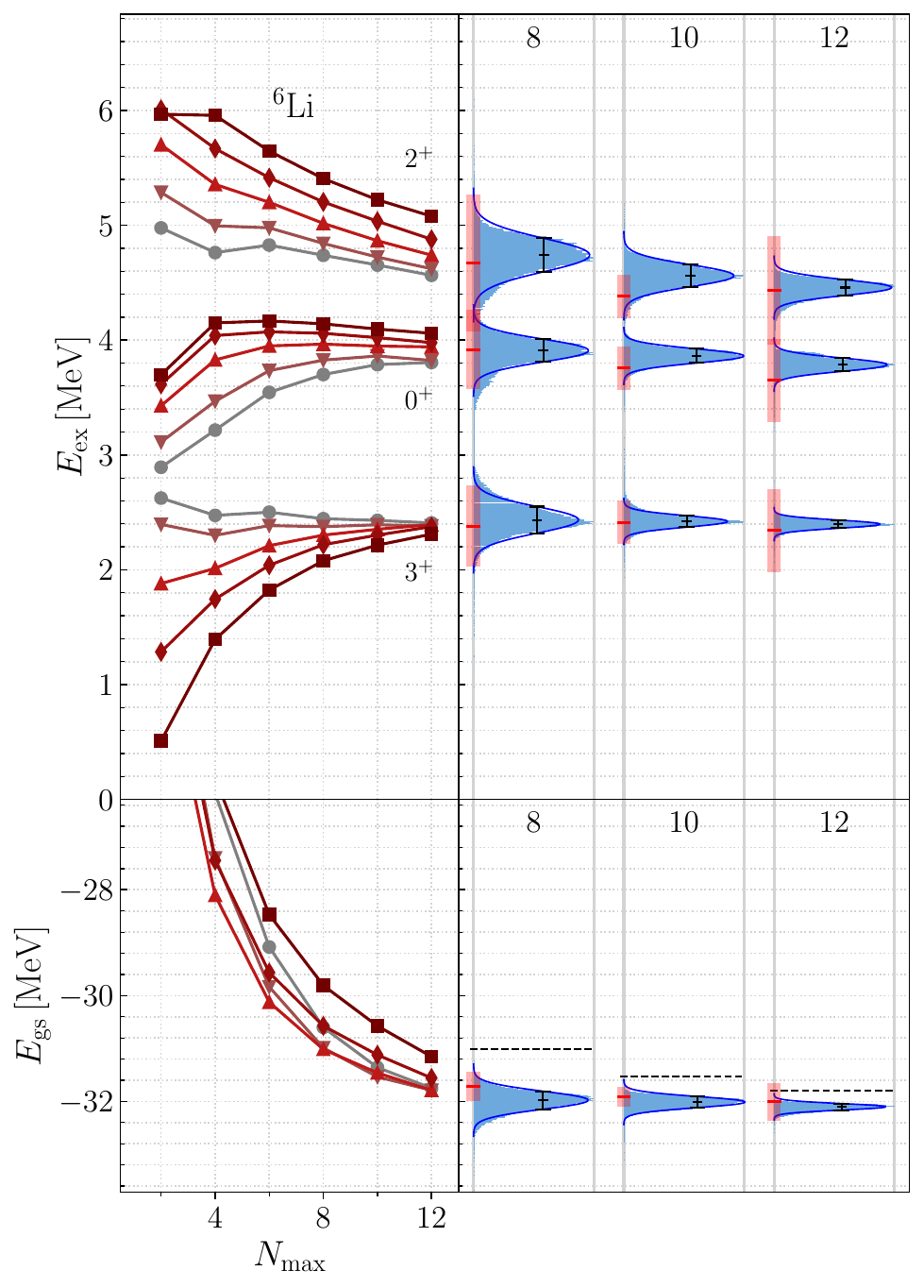}
    \caption{Input data and network predictions for the ground state and excitation energy of \elem{Li}{6}. The lower left hand panels shows the NCSM input data for the ground-state prediction, while the upper panel illustrated the convergence behavior of the exited states, with data for $\hbar\Omega=\qtylist{14;16;20;24;28}{MeV}$ (gray to red). The right hand panels show the corresponding network predictions, together with the classical extrapolation results and the variational minimum for the ground-state energy (dashed line).
    Note that the ANN predictions for the excitation energies are based on the sample-wise difference of absolute energy predictions. The corresponding convergence patterns in the top left panel are only shown for comparison.}
    \label{fig:E*_Li6}
\end{figure}
\begin{figure}[t]
    \includegraphics[width=1\columnwidth]{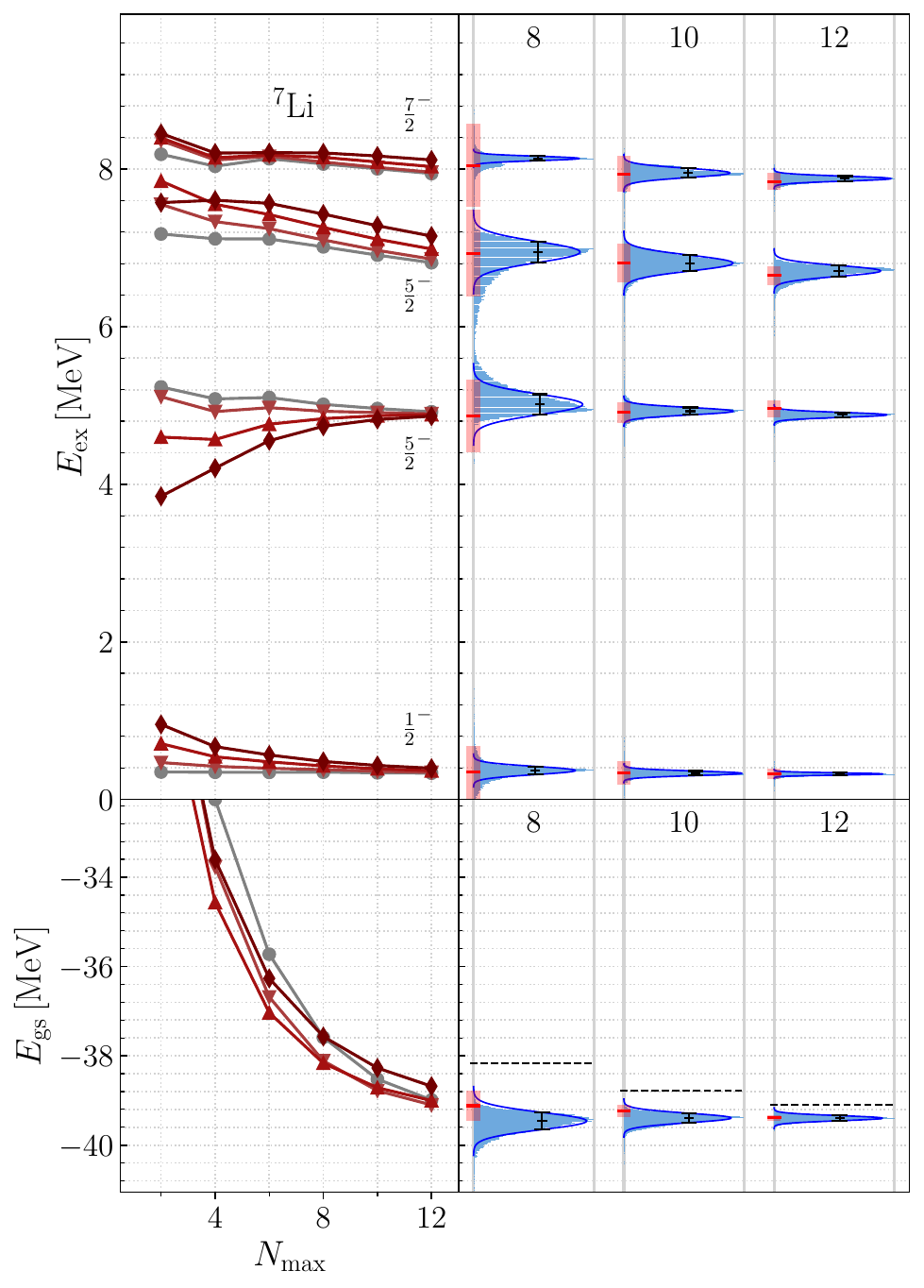}
    \caption{Input data and network predictions for the ground state and excitation energy of \elem{Li}{7} with $\hbar\Omega=\qtylist{14;16;20;24}{MeV}$ (gray to red), analogous to \cref{fig:E*_Li6}}
    \label{fig:E*_Li7}
\end{figure}
\begin{figure}[t]
    \includegraphics[width=1\columnwidth]{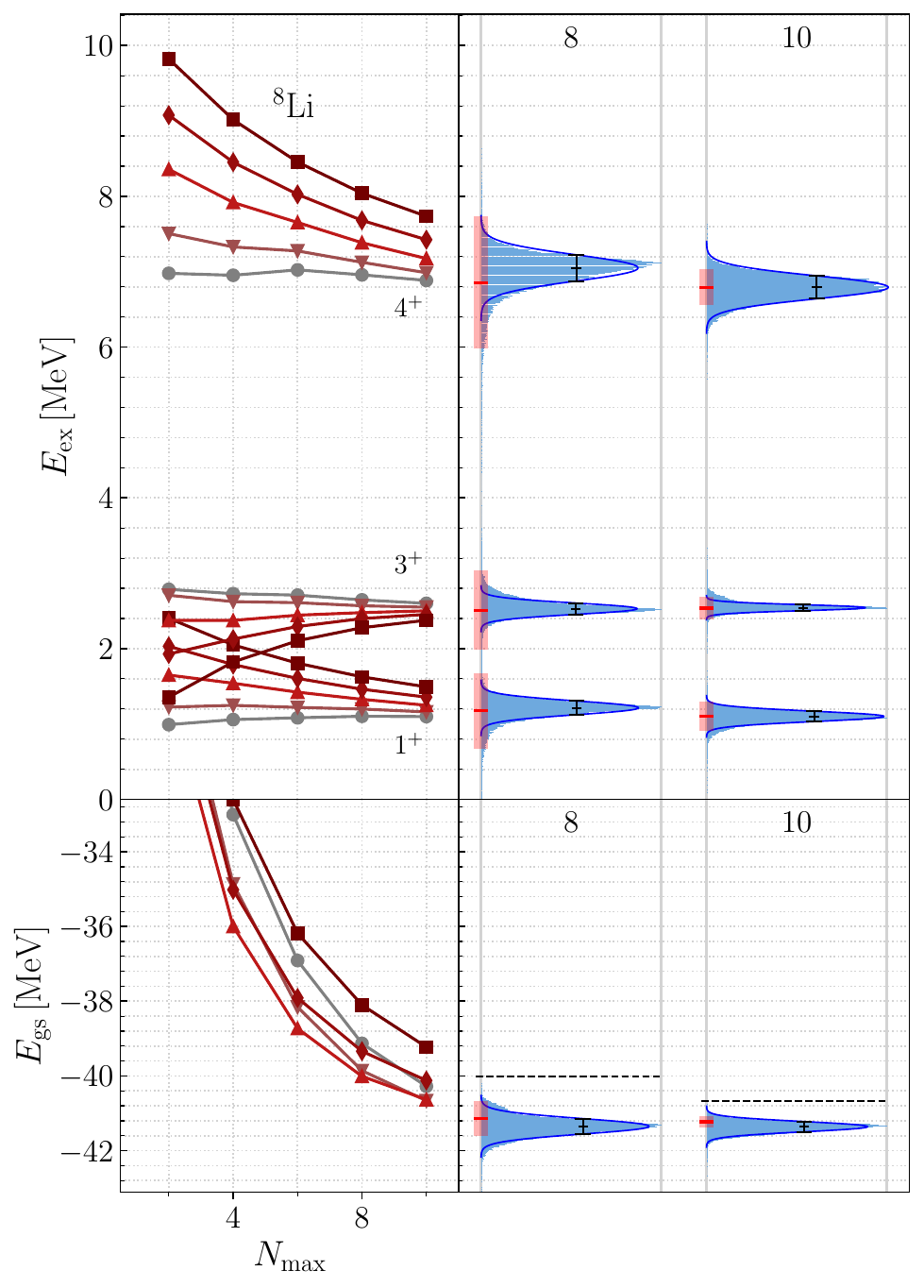}
    \caption{Input data and network predictions for the ground state and excitation energy of \elem{Li}{8} with $\hbar\Omega=\qtylist{14;16;20;24;28}{MeV}$ (gray to red), analogous to \cref{fig:E*_Li6}.}
    \label{fig:E*_Li8}
\end{figure}
\begin{figure}[t]
    \includegraphics[width=1\columnwidth]{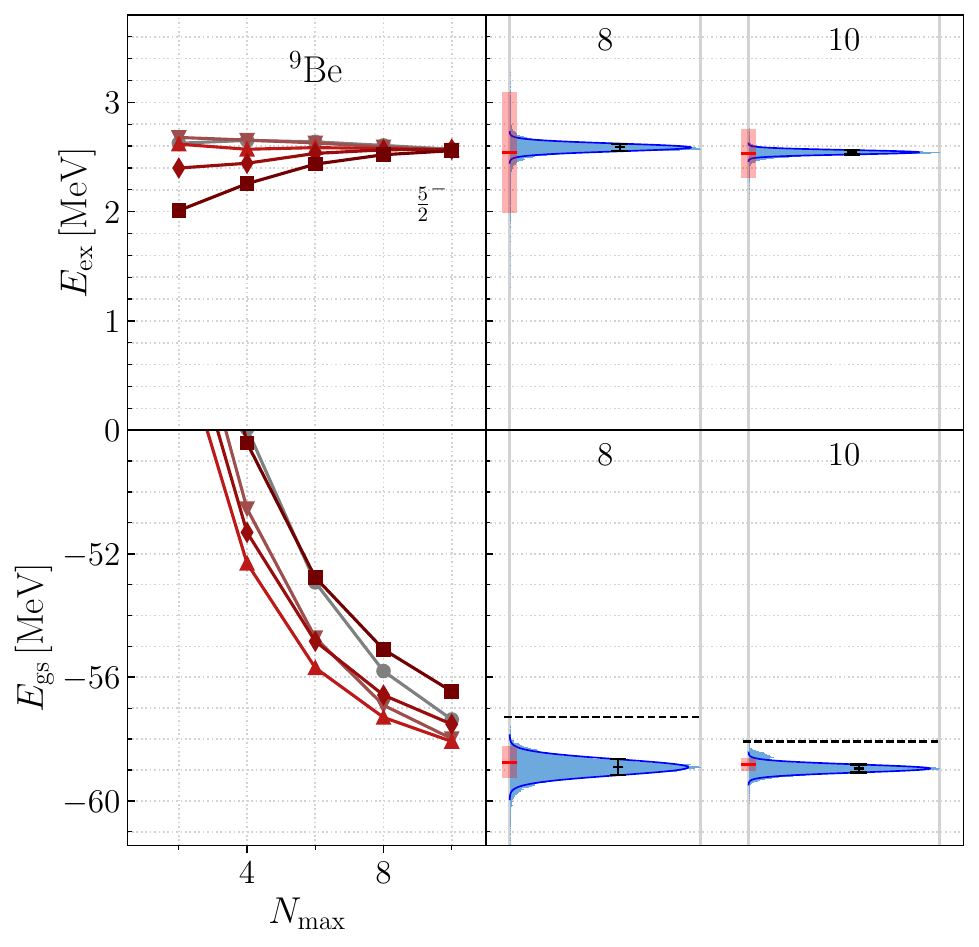}
    \caption{Input data and network predictions for the ground state and excitation energy of \elem{Be}{9} with $\hbar\Omega=\qtylist{14;16;20;24;28}{MeV}$ (gray to red), analogous to \cref{fig:E*_Li6}.}
    \label{fig:E*_Be9}
\end{figure}

In order to evaluate the performance of the networks, we will start with a look at ground-state energies of the few-body systems \elem{H}{2}, \elem{H}{3}, and \elem{He}{4}, which are also used in the train process.
It is important to note that all evaluations shown in this work use an entirely different family of interactions than the non-local chiral interaction used during training. The evaluation data are obtained with a semi-local NN+3N interaction derived from chiral EFT at $\NxLO{2}$ with a cutoff of $\Lambda=\qty{450}{MeV}$ \cite{ReKr18,LENPIC21}, SRG evolved with a flow parameter of $\alpha=\qty{0.08}{fm^4}$. This ensures maximum independence of training and evaluation data, even if the evaluation is performed for isotopes contained in the training data. The specific selection of HO frequencies for the NCSM evaluation data is given in the caption of the respective figure.

\Cref{fig:LightEgs} shows results for the few-body nuclei, the corresponding numerical values are summarized in \cref{tab:summary}.
As before, the left-hand panels show the raw evaluation data, while the right-hand panels show the distribution of the predictions together with a fitted Gaussian that defines the mean and standard deviation. The dotted lines indicate the variational minimum for $\Nmaxmax=\numlist{8;10;12}$ truncating the data fed into the networks. The results of traditional extrapolations at $\Nmax=\numlist{8;10;12}$ are given in red for comparison. The green line is the fully converged result obtained at very large $\Nmax$.

As expected, the networks show excellent agreement with the fully converged results and are well below the variational minimum except when the sequences are already well converged. The prediction using the smallest $\Nmaxmax=8$ is in all three cases almost exactly on the fully converged value, indicating that the networks can recognize the convergence pattern very well and give accurate predictions. The uncertainties of the predictions get systematically smaller with increasing $\Nmaxmax$, and all predictions are well within the preceding uncertainties indicating over-all consistency of the predictions and the uncertainty estimated.

Compared to the classical extrapolation, the networks are generally more accurate and precise in cases where the energy is not converged yet, which is the case for \elem{H}{2} and \elem{H}{3}, especially at smaller $\Nmaxmax$. Only the exceptionally well bound \elem{He}{4} nucleus converges so fast, that there is no advantage compared to a classical extrapolation. It should be noted that for \elem{H}{2} at $\Nmax=10$ the uncertainty estimation of the classical extrapolation breaks down (specifically the exponential fit for $\hbar\Omega=\SI{12}{MeV}$)
due to a deuteron-specific artifact in the convergence sequence, resulting in an unreasonably large uncertainty.

Looking at slightly heavier nuclei, we are not only interested in the ground-state energy, but also in excited-state energies or excitation energies.
The energy of excited states can be predicted in exactly the same way using the same pre-trained networks as for the ground-state energy extrapolation, feeding sequences of the converging excited state to the networks as inputs instead. The upper panel of \cref{fig:E*Example} shows the NCSM data for the $1^+$ ground and $3^+$ excited state of \elem{Li}{6}, with the corresponding network predictions on the right. There is no discernible difference in the performance of the networks for the two states, for both states the predictions are very stable and consistent with each other. This is possible since there is no qualitative difference in the convergence behavior of bound excited states compared to the ground state.

To give predictions for excitation energies with our networks, we evaluate both the ground and the excited-state energy separately, and subsequently subtract the predictions sample-wise, i.e., we compute the difference between the predicted ground-state energy and the predicted excited-state energy of for each sample consisting of $X=3$ different HO frequencies and $L=4$ subsequent model space truncations $N_{\max}$. The resulting distribution tends to be approximately normal-shaped, so that we can once again simply fit a Gaussian to obtain the prediction for the excitation energy and its uncertainty from the mean and standard deviation the Gaussian.

The lower panels of \cref{fig:E*Example} show the result of the above mentioned procedure for the $3^+$ excitation energy on the right-hand side. The excitation energy sequences in the left-hand panel are only shown to give an impression of the convergence of the actual observable, as they are not used as input data for the networks.
We observe that the excitation energy predictions are remarkably stable for the different $\Nmaxmax$, given that this observable is a lot more challenging to predict, as slight variations in the prediction of the absolute energies has larger effects on the excitation energy.
For this reason, we use the aforementioned sample-wise subtraction method instead of just taking the difference of the whole distributions. This way, we can make exploit of correlations in the convergence pattern of different states for the same HO frequency. Typically, this improves the prediction and leads to smaller uncertainties.

\Cref{fig:E*_Li6,fig:E*_Li7,fig:E*_Li8,fig:E*_Be9} as well as \cref{tab:summary} give a comprehensive overview of the performance of the networks for ground-state energies and the lowest few excitation energies of \elem{Li}{6}, \elem{Li}{7}, \elem{Li}{8}, and \elem{Be}{9}. These NCSM calculations for p-shell nuclei have been discussed in detail in Ref.~\cite{MaLe23} and we reuse these NCSM outputs here to benchmark the performance of the networks \cite{MaPriv}. In these figures, the ground-state energy is shown in the lower panel, while the excitation energy results are in the upper panel. Additionally, classical extrapolations for both observables are depicted as a red bar, with the corresponding uncertainty as a red band around it.

One major observation we want to highlight is that the ANN predictions for the ground-state energies are remarkably stable with increasing model-space sizes and generally significantly outperform the classical extrapolations at $\Nmaxmax=8$ across all considered nuclei when compared with the values obtained for large $\Nmaxmax$. The classical extrapolations are more susceptible to a systematic downwards trend with increasing model-space size, which has its roots in the fact that the convergence of ground-state energies is not quite exponential. The networks do not suffer from such a systematic, which facilitates a robust and accurate prediction of the converged energy in smaller model spaces.

Excitation energies are slightly more difficult, since the correlation between the convergence of the ground and the excited state are relevant. Furthermore, excited-state energies, especially for more loosely bound states, converge slower than the ground-state energy.
Together, this makes the prediction of the converged value much more challenging, especially at $\Nmaxmax=8$.
Nevertheless, the ANN results for excitation energies are generally very robust, in particular for the lowest excited states.
Higher-lying, more loosely bound excited states like the $2^+$ state in \elem{Li}{6} or the $\tfrac{7}{2}^-$ state in \elem{Li}{7} require $\Nmaxmax=10$ to give accurate results. However, in these cases, the convergence behavior of the sequences changes significantly within the lower $\Nmax$ data points, so a slight influence on the ANN predictions is expected. For this reason, we recommend checking for the robustness of the predictions by comparing with predictions for other $\Nmaxmax$ if available.

\section{Results for Radii}
\begin{figure}[t]
    \includegraphics[width=1\columnwidth]{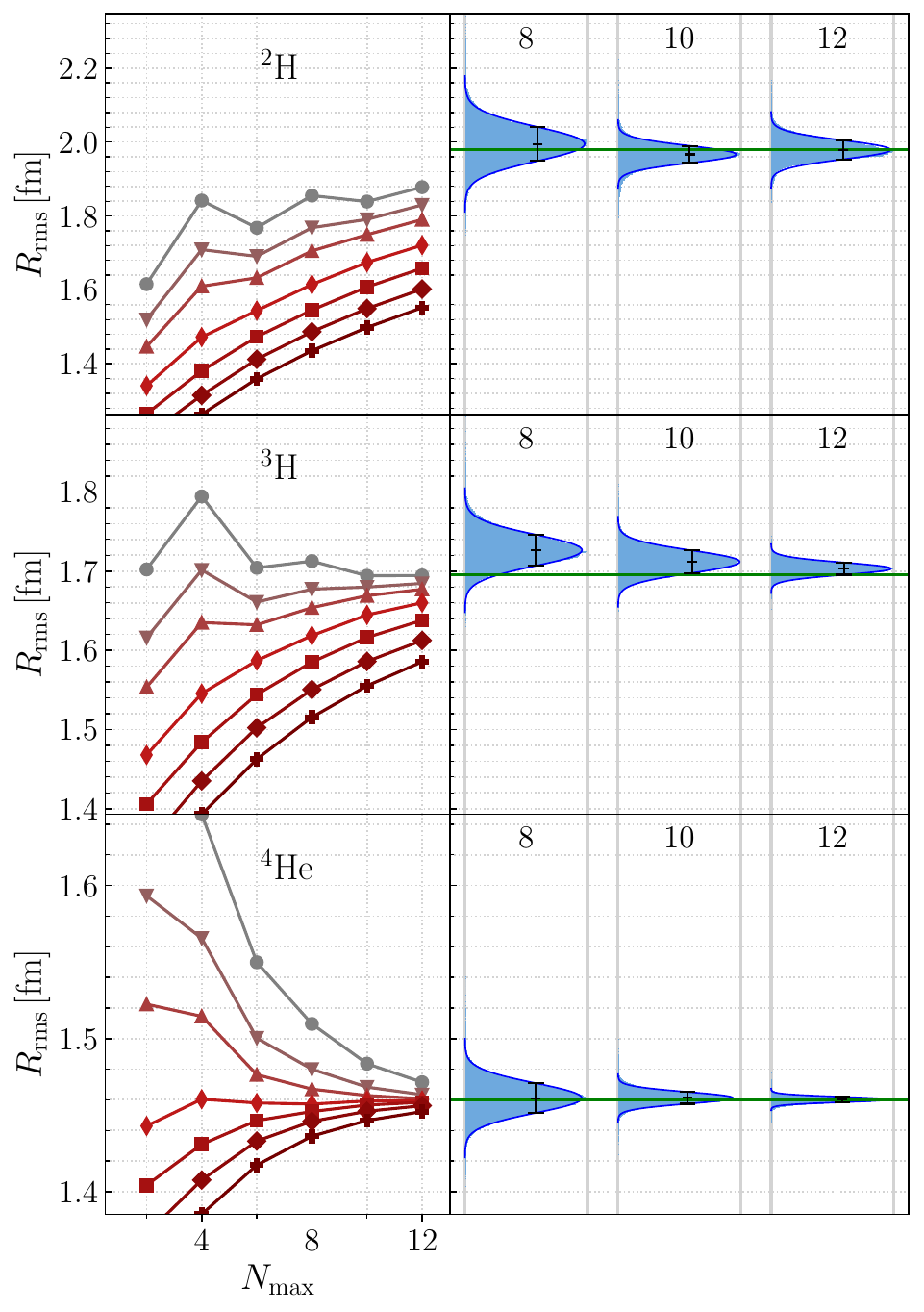}
    \caption{Mass rms radii input data and resulting network predictions for few-body systems. The left hand panels show the NCSM data for $\hbar\Omega=\qtylist{12;14;16;20;24;28;32}{MeV}$ (gray to red), while the right hand panels show the ANN predictions for $\Nmaxmax=\numlist{8;10;12}$, with the green line indicating the fully converged value obtained at a large $\Nmax$.}
    \label{fig:R_light}
\end{figure}
\begin{figure}[t]
    \includegraphics[width=1\columnwidth]{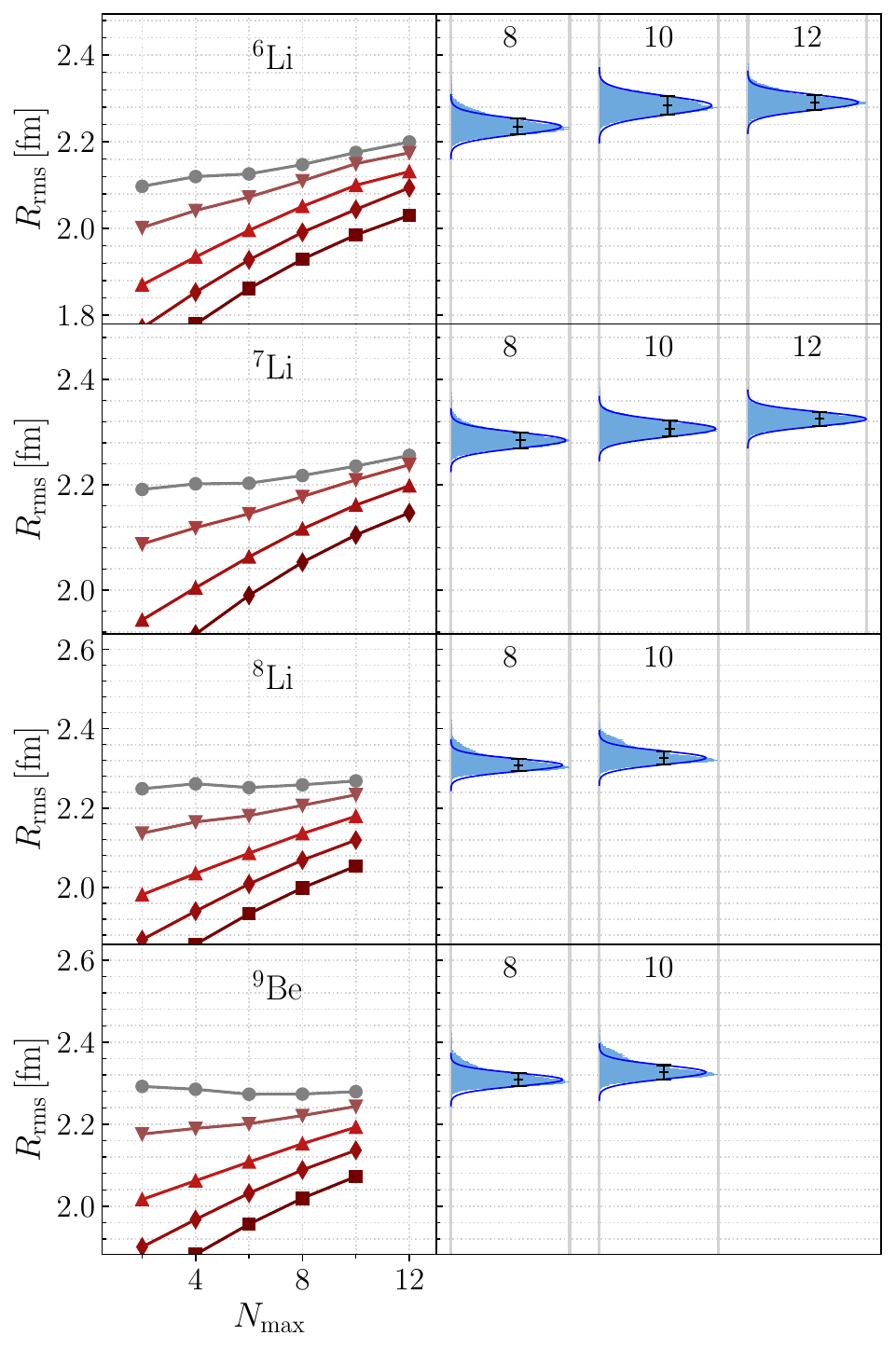}
    \caption{Mass rms radius input data and network predictions for light p-shell nuclei with $\hbar\Omega=\qtylist{14;16;20;24;28}{MeV}$ (gray to red, \elem{Li}{7} without $\hbar\Omega=\qty{28}{MeV}$), analogous to \cref{fig:R_light}.}
    \label{fig:R_heavy}
\end{figure}
\begin{table}\footnotesize
    \begin{tabular}{l l S[table-format=4.7] S[table-format=3.7,table-align-text-after = false] S[table-format=3.7] S[table-format=3.3]}
        \hline\hline
        & & {$\mathcal{N}_\mathrm{max}=8$} & {10} & {12} & {conv.}\\
        \hline\\[-1em]
        \multicolumn{5}{c}{\elem{H}{2}} \\
        \hline
        $E_\text{gs}$  & ANN                 & -2.195(47) & -2.223(27) & -2.195(14) & -2.200 \\
        $E_\text{gs}$  & extr.           & -2.013(106) & -2.147{(*)} & -2.183(106) & -2.200\\
        $R_\text{rms}$  & ANN                  &  1.995(46)  &  1.967(23)  &  1.979(26)  &  1.979\\[0.3em]
        \multicolumn{5}{c}{\elem{H}{3}} \\
        \hline
        $E_\text{gs}$  & ANN                 & -8.482(65)  &  -8.509(33)  &  -8.487(12)  &  -8.481 \\
        $E_\text{gs}$  & extr.           & -8.434(110) &  -8.480(109) &  -8.473(24)  &  -8.481 \\
        $R_\text{rms}$  & ANN                  &  1.726(20)  &  1.712(14)   &  1.703(8)  &  1.696\\[0.3em]
        \multicolumn{5}{c}{\elem{He}{4}} \\
        \hline
        $E_\text{gs}$  & ANN                 & -28.513(71)  & -28.526(25)  & -28.526(8)   &  -28.524\\
        $E_\text{gs}$  & extr.           & -28.503(48)  & -28.528(24)  & -28.523(9)   &  -28.524 \\
        $R_\text{rms}$  & ANN                  &  1.461(10)  &   1.461(4)   &   1.460(2)   &    1.460\\[0.3em]
        \multicolumn{5}{c}{\elem{Li}{6}} \\
        \hline
        $E_\text{gs}$  & ANN         & -31.98(17) & -32.01(11) & -32.10(6)  & {--} \\
        $E_\text{gs}$  & extr.     & -31.72(27) & -31.91(19) & -32.01(36)  & {--} \\
        $E_\text{ex}(3^+)$  & ANN     &   2.44(12) &   2.42(5)  &   2.40(4)  & {--} \\
        $E_\text{ex}(3^+)$  & extr. &   2.38(35) &   2.41(19) &   2.34(36) & {--} \\
        $E_\text{ex}(0^+)$  & ANN     &   3.91(10)  &   3.86(6)  &   3.79(6)  & {--} \\
        $E_\text{ex}(0^+)$  & extr. &   3.92(35) &   3.76(19) &   3.65(36) & {--} \\
        $E_\text{ex}(2^+)$  & ANN     &   4.74(15) &   4.56(10)  &   4.46(7)  & {--} \\
        $E_\text{ex}(2^+)$  & extr. &   4.67(59) &   4.38(19) &   4.44(47) & {--} \\
        $R_\text{rms}$  & ANN	       &   2.235(18)  &   2.284(22)  &   2.291(18)  & {--} \\[0.3em]
        \multicolumn{5}{c}{\elem{Li}{7}} \\
        \hline
        $E_\text{gs}$  & ANN         & -39.46(19) & -39.39(11) & -39.40(6)  & {--} \\
        $E_\text{gs}$  & extr.     & -39.12(34) & -39.24(13) & -39.39(7)  & {--} \\
        $E_\text{ex}(1/2^-)$  & ANN     &   0.37(5)  &   0.33(3)  &   0.32(1)  & {--} \\
        $E_\text{ex}(1/2^-)$  & extr. &   0.34(34) &   0.33(15) &   0.32(7) & {--} \\
        $E_\text{ex}(5/2^-)$  & ANN     &   5.01(13) &   4.93(5)  &   4.88(3)  & {--} \\
        $E_\text{ex}(5/2^-)$  & extr. &   4.86(46) &   4.91(13) &   4.96(11) & {--} \\
        $E_\text{ex}(5/2^-)$  & ANN     &   6.95(13) &   6.81(10) &   6.71(7)  & {--} \\
        $E_\text{ex}(5/2^-)$  & extr. &   6.93(55) &   6.81(25) &   6.65(12) & {--} \\
        $E_\text{ex}(7/2^-)$  & ANN     &   8.14(3)  &   7.95(6)  &   7.88(3)  & {--} \\
        $E_\text{ex}(7/2^-)$  & extr. &   8.05(53) &   7.94(23) &   7.84(11) & {--} \\
        $R_\text{rms}$  & ANN	       &   2.285(15)  &   2.307(15)  &   2.325(14)  & {--} \\[0.3em]
        \multicolumn{5}{c}{\elem{Li}{8}} \\
        \hline
        $E_\text{gs}$  & ANN         & -41.35(21) & -41.36(14)  & {--}  & {--} \\
        $E_\text{gs}$  & extr.     & -41.14(47) & -41.23(16)  & {--}  & {--} \\
        $E_\text{ex}(1^+)$  & ANN     &   1.22(9)  &   1.10(7)  & {--}  & {--} \\
        $E_\text{ex}(1^+)$  & extr. &   1.18(50) &   1.11(19)  & {--} & {--} \\
        $E_\text{ex}(3^+)$  & ANN     &   2.53(8)  &   2.54(4)  & {--}  & {--} \\
        $E_\text{ex}(3^+)$  & extr. &   2.51(52) &   2.54(16)  & {--} & {--} \\
        $E_\text{ex}(4^+)$  & ANN     &   7.05(17) &   6.79(15)  & {--}  & {--} \\
        $E_\text{ex}(4^+)$  & extr. &   6.86(88) &   6.80(24)  & {--} & {--} \\
        $R_\text{rms}$  & ANN	       &   2.308(16)  &   2.327(17)  & {--}  & {--} \\[0.3em]
        \multicolumn{5}{c}{\elem{Be}{9}} \\
        \hline
        $E_\text{gs}$  & ANN         & -58.91(26) & -58.95(13) & {--}  & {--} \\
        $E_\text{gs}$  & extr.     & -58.74(52) & -58.82(21) & {--}  & {--} \\
        $E_\text{ex}(5/2^-)$  & ANN     &   2.59(4)  &   2.54(2)  & {--}  & {--} \\
        $E_\text{ex}(5/2^-)$  & extr. &   2.54(55) &   2.53(23) & {--} & {--} \\
        $R_\text{rms}$  & ANN	       &   2.309(14)  &   2.324(13)  & {--}  & {--} \\
\hline\hline
    \end{tabular}
    \caption{ANN predictions of ground state and excitation energies [MeV] and the mass rms radius [fm] for all discussed nuclei. Classical extrapolation results for energies are given as a comparison. Where available, the fully converged result is given as well. Uncertainties denoted as (*) are unreasonably large due to a breakdown of the classical extrapolation method.}
     \label{tab:summary}
\end{table}
As already mentioned, the extrapolation of NCSM sequences for rms radii is very challenging, as their convergence behavior is far less constrained than for the ground-state energy. As a result a larger variety of convergence patterns can be observed and the networks have to learn to cope with the different patterns. Obviously, we have to train a new set of networks to pick up the convergence patterns inherent to the radius. We use the same network topology and input mode (MINMAX) as for the ground-state energy training, but train on data for mass rms radii instead. The evaluation step is performed analogous to the ground-state energies.

The application of the radius networks to \elem{H}{2}, \elem{H}{3}, and \elem{He}{4} is illustrated in \cref{fig:R_light} and the final ANN predictions are summarized in \cref{tab:summary}. The NCSM data for the mass rms radius, which is fed into the networks, is shown in the left-hand panels, while the ANN predictions are depicted on the right side, together with the fully converged values (green lines). As before, we use the semi-local NN+3N interaction at $\NxLO{2}$ with $\Lambda=\qty{450}{MeV}$ \cite{ReKr18,LENPIC21} and $\alpha=\qty{0.08}{fm^4}$ for this evaluation, which is not included in the training set.

We first observe that the convergence patterns for the radii are really very different from the ground-state energies. The radius is not constrained via the variational principle, which is evident from the fact that convergence is not monotonous. Depending on the selection of HO frequencies, whole sequences can converge from or from below. The convergence itself is also significantly slower compared to energies, and some sequences show a jagged pattern, particularly at $\Nmax$ values.

Despite of these characteristics, the ANN predictions for the \elem{H}{2} and \elem{He}{4} mass rms radius are in very good agreement with the fully converged value, even at $\Nmaxmax=8$. These two nuclei also demonstrate the two extremes of possible convergence patterns: With a selection of HO frequencies around the variational minimum for the ground-state energy, \elem{He}{4} exhibits a very symmetric convergence, with some sequences converging from above and some from below. \elem{H}{2} on the other hand exhibits convergence exclusively from below, which is the typical case for most nuclei we looked at. This reflects the distribution of target values for all training samples mentioned earlier, and depicted in \cref{fig:MINMAX_target_distribution}. The predictions for \elem{H}{3} show a light drift, particularly the $\Nmaxmax=8$ is slightly above the exact radius, which might be related to the strong zig-zag pattern for small $N_{\max}$ values in the input data.

To assess the performance of the networks beyond of the few-body systems used for training, we take a look at four p-shell nuclei. \Cref{fig:R_heavy} shows input data and ANN predictions for \elem{Li}{6}, \elem{Li}{7}, \elem{Li}{8}, and \elem{Be}{9}. The final predictions are summarized in \cref{tab:summary}. These data again are extracted from the NCSM calculations performed for Ref.~\cite{MaLe23}, which focused entirely on energies. Therefore, the selection of HO frequencies available is optimized for the description of energies and not for rms radii. As a result, the sequences predominantly converge from below and, particularly for \elem{Li}{6} and \elem{Li}{7}, do not cover the range around the optimum frequency for the radius.

Despite these deficiencies in the evaluation data, the predictions are very stable across the board. For \elem{Li}{6} and \elem{Li}{7} a slight trend towards larger radii, which is compatible with the depicted uncertainties, can be observed with increasing $\Nmaxmax$. This shift gets smaller for larger model spaces, and is less pronounced for \elem{Li}{8} and \elem{Be}{9}, where the chosen HO frequencies include more optimal values producing a flat convergence curve.
Therefore, we expect that an optimized selection of HO frequencies in the NCSM calculations, leading to a more symmetric convergence pattern, will improve the predictions and alleviate the observed slight upwards trend.
Another way to tackle such a systematic might be to restrict the training data set to sequences that converge exclusively from below, thereby, obtaining networks that are specialized for this situation.
However, even with a non-optimized frequency selection and the general set of training data, the ANNs provide robust predictions for the converged rms radii with realistic data-driven uncertainties based on a rather limited set of evaluation data.

\section{Conclusions}

We have extended our ANN architecture to high-precision predictions of ground-state and excited-state energies and radii from non-converged NCSM calculations. The core idea of the framework is to learn convergence patterns from a large pool of training data obtained from few-body systems such as \elem{H}{2}, \elem{H}{3}, and \elem{He}{4} with a range of different interactions, and apply the trained networks to NCSM data for a larger range of nuclei. This includes a statistical evaluation scheme, where many networks are trained and evaluated to obtain a distribution of predictions that allows for the extraction of statistically meaningful uncertainties. With the new MINMAX input mode we are able to extend this capability to radii as well, a much less constrained observable compared to energies. Additionally, we find improvements in accuracy coupled with more realistic uncertainties in the case of energies when compared with our previous generation of networks.

Based on these advances we are convinced that the ANN architecture is capable of handling other observable as well, especially  electromagnetic moments and transition strengths. For electric quadrupole observables we expect a correlation to the rms radius, which can be exploited in the evaluation and construction of the networks. Of course, for the training of new ANN for such observables we need a sufficient amount of training data in very light nuclei. This can become an issue, since many electromagnetic observables, particularly transition strengths, do not naturally appear in bound few-nucleon system. To remedy this issue, we are currently looking into synthetic nuclei which emerge from modified Hamiltonians. We can, for example, enhance the interaction to create additional bound states in these small systems, making more complex observables accessible in few-body systems and to create a large library of training data.

\begin{acknowledgments}

We thank Pieter Maris for providing us with the raw NCSM output for the calculations presented in Refs. \cite{LENPIC21,MaLe23}, which were performed at the Argonne Leadership Computing Facility (ALCF), a US Department of Energy Office of Science user facility, supported under Contract No. DE-AC02-06CH11357, using computing resources provided  under the INCITE award `Nuclear Structure and Nuclear Reactions' from the US Department of Energy, Office of Advanced Scientific Computing Research.
All other numerical calculations have been performed on the LICHTENBERG II cluster at the computing center of the TU Darmstadt.
This work is supported by the Deutsche For\-schungs\-ge\-mein\-schaft (DFG, German Research\linebreak Foundation) through the DFG Sonderforschungsbereich SFB 1245 (Project-ID 279384907) and the BMBF through Verbundprojekt 05P2021 (ErUM-FSP T07, Contract No. 05P21RDFNB).

\end{acknowledgments}

\bibliography{bib_arXiv_v2}

\begin{thebibliography}{35}%
\makeatletter
\providecommand \@ifxundefined [1]{%
 \@ifx{#1\undefined}
}%
\providecommand \@ifnum [1]{%
 \ifnum #1\expandafter \@firstoftwo
 \else \expandafter \@secondoftwo
 \fi
}%
\providecommand \@ifx [1]{%
 \ifx #1\expandafter \@firstoftwo
 \else \expandafter \@secondoftwo
 \fi
}%
\providecommand \natexlab [1]{#1}%
\providecommand \enquote  [1]{``#1''}%
\providecommand \bibnamefont  [1]{#1}%
\providecommand \bibfnamefont [1]{#1}%
\providecommand \citenamefont [1]{#1}%
\providecommand \href@noop [0]{\@secondoftwo}%
\providecommand \href [0]{\begingroup \@sanitize@url \@href}%
\providecommand \@href[1]{\@@startlink{#1}\@@href}%
\providecommand \@@href[1]{\endgroup#1\@@endlink}%
\providecommand \@sanitize@url [0]{\catcode `\\12\catcode `\$12\catcode
  `\&12\catcode `\#12\catcode `\^12\catcode `\_12\catcode `\%12\relax}%
\providecommand \@@startlink[1]{}%
\providecommand \@@endlink[0]{}%
\providecommand \url  [0]{\begingroup\@sanitize@url \@url }%
\providecommand \@url [1]{\endgroup\@href {#1}{\urlprefix }}%
\providecommand \urlprefix  [0]{URL }%
\providecommand \Eprint [0]{\href }%
\providecommand \doibase [0]{https://doi.org/}%
\providecommand \selectlanguage [0]{\@gobble}%
\providecommand \bibinfo  [0]{\@secondoftwo}%
\providecommand \bibfield  [0]{\@secondoftwo}%
\providecommand \translation [1]{[#1]}%
\providecommand \BibitemOpen [0]{}%
\providecommand \bibitemStop [0]{}%
\providecommand \bibitemNoStop [0]{.\EOS\space}%
\providecommand \EOS [0]{\spacefactor3000\relax}%
\providecommand \BibitemShut  [1]{\csname bibitem#1\endcsname}%
\let\auto@bib@innerbib\@empty
\bibitem [{\citenamefont {Barrett}\ \emph {et~al.}(2013)\citenamefont
  {Barrett}, \citenamefont {Navr{\'a}til},\ and\ \citenamefont
  {Vary}}]{BaNa13}%
  \BibitemOpen
  \bibfield  {author} {\bibinfo {author} {\bibfnamefont {B.~R.}\ \bibnamefont
  {Barrett}}, \bibinfo {author} {\bibfnamefont {P.}~\bibnamefont
  {Navr{\'a}til}},\ and\ \bibinfo {author} {\bibfnamefont {J.~P.}\ \bibnamefont
  {Vary}},\ }\bibfield  {title} {\bibinfo {title} {Ab initio no core shell
  model},\ }\href {https://doi.org/10.1016/j.ppnp.2012.10.003} {\bibfield
  {journal} {\bibinfo  {journal} {Prog. Part. Nucl. Phys.}\ }\textbf {\bibinfo
  {volume} {69}},\ \bibinfo {pages} {131} (\bibinfo {year} {2013})}\BibitemShut
  {NoStop}%
\bibitem [{\citenamefont {Navr{\'a}til}\ \emph {et~al.}(2009)\citenamefont
  {Navr{\'a}til}, \citenamefont {Quaglioni}, \citenamefont {Stetcu},\ and\
  \citenamefont {Barrett}}]{NaQu09}%
  \BibitemOpen
  \bibfield  {author} {\bibinfo {author} {\bibfnamefont {P.}~\bibnamefont
  {Navr{\'a}til}}, \bibinfo {author} {\bibfnamefont {S.}~\bibnamefont
  {Quaglioni}}, \bibinfo {author} {\bibfnamefont {I.}~\bibnamefont {Stetcu}},\
  and\ \bibinfo {author} {\bibfnamefont {B.~R.}\ \bibnamefont {Barrett}},\
  }\bibfield  {title} {\bibinfo {title} {Recent developments in no-core
  shell-model calculations},\ }\href
  {https://doi.org/10.1088/0954-3899/36/8/083101} {\bibfield  {journal}
  {\bibinfo  {journal} {J. Phys. G: Nucl. Part. Phys.}\ }\textbf {\bibinfo
  {volume} {36}},\ \bibinfo {pages} {083101} (\bibinfo {year}
  {2009})}\BibitemShut {NoStop}%
\bibitem [{\citenamefont {Roth}(2009)}]{Roth09}%
  \BibitemOpen
  \bibfield  {author} {\bibinfo {author} {\bibfnamefont {R.}~\bibnamefont
  {Roth}},\ }\bibfield  {title} {\bibinfo {title} {Importance truncation for
  large-scale configuration interaction approaches},\ }\href
  {https://doi.org/10.1103/PhysRevC.79.064324} {\bibfield  {journal} {\bibinfo
  {journal} {Phys. Rev. C}\ }\textbf {\bibinfo {volume} {79}},\ \bibinfo
  {pages} {064324} (\bibinfo {year} {2009})}\BibitemShut {NoStop}%
\bibitem [{\citenamefont {Zheng}\ \emph {et~al.}(1993)\citenamefont {Zheng},
  \citenamefont {Barrett}, \citenamefont {Jaqua}, \citenamefont {Vary},\ and\
  \citenamefont {McCarthy}}]{ZhBa93}%
  \BibitemOpen
  \bibfield  {author} {\bibinfo {author} {\bibfnamefont {D.~C.}\ \bibnamefont
  {Zheng}}, \bibinfo {author} {\bibfnamefont {B.~R.}\ \bibnamefont {Barrett}},
  \bibinfo {author} {\bibfnamefont {L.}~\bibnamefont {Jaqua}}, \bibinfo
  {author} {\bibfnamefont {J.~P.}\ \bibnamefont {Vary}},\ and\ \bibinfo
  {author} {\bibfnamefont {R.~J.}\ \bibnamefont {McCarthy}},\ }\bibfield
  {title} {\bibinfo {title} {Microscopic calculations of the spectra of light
  nuclei},\ }\href {https://doi.org/10.1103/PhysRevC.48.1083} {\bibfield
  {journal} {\bibinfo  {journal} {Phys. Rev. C}\ }\textbf {\bibinfo {volume}
  {48}},\ \bibinfo {pages} {1083} (\bibinfo {year} {1993})}\BibitemShut
  {NoStop}%
\bibitem [{\citenamefont {Kowalski}\ \emph {et~al.}(2004)\citenamefont
  {Kowalski}, \citenamefont {Dean}, \citenamefont {{Hjorth-Jensen}},
  \citenamefont {Papenbrock},\ and\ \citenamefont {Piecuch}}]{KoDe04}%
  \BibitemOpen
  \bibfield  {author} {\bibinfo {author} {\bibfnamefont {K.}~\bibnamefont
  {Kowalski}}, \bibinfo {author} {\bibfnamefont {D.~J.}\ \bibnamefont {Dean}},
  \bibinfo {author} {\bibfnamefont {M.}~\bibnamefont {{Hjorth-Jensen}}},
  \bibinfo {author} {\bibfnamefont {T.}~\bibnamefont {Papenbrock}},\ and\
  \bibinfo {author} {\bibfnamefont {P.}~\bibnamefont {Piecuch}},\ }\bibfield
  {title} {\bibinfo {title} {Coupled {{Cluster Calculations}} of {{Ground}} and
  {{Excited States}} of {{Nuclei}}},\ }\href
  {https://doi.org/10.1103/PhysRevLett.92.132501} {\bibfield  {journal}
  {\bibinfo  {journal} {Phys. Rev. Lett.}\ }\textbf {\bibinfo {volume} {92}},\
  \bibinfo {pages} {132501} (\bibinfo {year} {2004})}\BibitemShut {NoStop}%
\bibitem [{\citenamefont {Dickhoff}\ and\ \citenamefont
  {Barbieri}(2004)}]{DiBa04}%
  \BibitemOpen
  \bibfield  {author} {\bibinfo {author} {\bibfnamefont {W.}~\bibnamefont
  {Dickhoff}}\ and\ \bibinfo {author} {\bibfnamefont {C.}~\bibnamefont
  {Barbieri}},\ }\bibfield  {title} {\bibinfo {title} {Self-consistent
  {{Green}}'s function method for nuclei and nuclear matter},\ }\href
  {https://doi.org/10.1016/j.ppnp.2004.02.038} {\bibfield  {journal} {\bibinfo
  {journal} {Prog. Part. Nucl. Phys.}\ }\textbf {\bibinfo {volume} {52}},\
  \bibinfo {pages} {377} (\bibinfo {year} {2004})}\BibitemShut {NoStop}%
\bibitem [{\citenamefont {Hergert}\ \emph {et~al.}(2016)\citenamefont
  {Hergert}, \citenamefont {Bogner}, \citenamefont {Morris}, \citenamefont
  {Schwenk},\ and\ \citenamefont {Tsukiyama}}]{HeBo16}%
  \BibitemOpen
  \bibfield  {author} {\bibinfo {author} {\bibfnamefont {H.}~\bibnamefont
  {Hergert}}, \bibinfo {author} {\bibfnamefont {S.}~\bibnamefont {Bogner}},
  \bibinfo {author} {\bibfnamefont {T.}~\bibnamefont {Morris}}, \bibinfo
  {author} {\bibfnamefont {A.}~\bibnamefont {Schwenk}},\ and\ \bibinfo {author}
  {\bibfnamefont {K.}~\bibnamefont {Tsukiyama}},\ }\bibfield  {title} {\bibinfo
  {title} {The {{In-Medium Similarity Renormalization Group}}: {{A}} novel ab
  initio method for nuclei},\ }\href
  {https://doi.org/10.1016/j.physrep.2015.12.007} {\bibfield  {journal}
  {\bibinfo  {journal} {Phys. Rep.}\ }\textbf {\bibinfo {volume} {621}},\
  \bibinfo {pages} {165} (\bibinfo {year} {2016})}\BibitemShut {NoStop}%
\bibitem [{\citenamefont {Carlson}\ \emph {et~al.}(2015)\citenamefont
  {Carlson}, \citenamefont {Gandolfi}, \citenamefont {Pederiva}, \citenamefont
  {Pieper}, \citenamefont {Schiavilla}, \citenamefont {Schmidt},\ and\
  \citenamefont {Wiringa}}]{CaGa15}%
  \BibitemOpen
  \bibfield  {author} {\bibinfo {author} {\bibfnamefont {J.}~\bibnamefont
  {Carlson}}, \bibinfo {author} {\bibfnamefont {S.}~\bibnamefont {Gandolfi}},
  \bibinfo {author} {\bibfnamefont {F.}~\bibnamefont {Pederiva}}, \bibinfo
  {author} {\bibfnamefont {S.~C.}\ \bibnamefont {Pieper}}, \bibinfo {author}
  {\bibfnamefont {R.}~\bibnamefont {Schiavilla}}, \bibinfo {author}
  {\bibfnamefont {K.~E.}\ \bibnamefont {Schmidt}},\ and\ \bibinfo {author}
  {\bibfnamefont {R.~B.}\ \bibnamefont {Wiringa}},\ }\bibfield  {title}
  {\bibinfo {title} {Quantum {{Monte Carlo}} methods for nuclear physics},\
  }\href {https://doi.org/10.1103/RevModPhys.87.1067} {\bibfield  {journal}
  {\bibinfo  {journal} {Rev. Mod. Phys.}\ }\textbf {\bibinfo {volume} {87}},\
  \bibinfo {pages} {1067} (\bibinfo {year} {2015})}\BibitemShut {NoStop}%
\bibitem [{\citenamefont {Clark}\ \emph {et~al.}(1999)\citenamefont {Clark},
  \citenamefont {Lindenau},\ and\ \citenamefont {Ristig}}]{ClLi99}%
  \BibitemOpen
  \bibinfo {editor} {\bibfnamefont {J.~W.}\ \bibnamefont {Clark}}, \bibinfo
  {editor} {\bibfnamefont {T.}~\bibnamefont {Lindenau}},\ and\ \bibinfo
  {editor} {\bibfnamefont {M.~L.}\ \bibnamefont {Ristig}},\ eds.,\ \href
  {https://doi.org/10.1007/BFb0104276} {\emph {\bibinfo {title} {Scientific
  {{Applications}} of {{Neural Nets}}}}},\ \bibinfo {series} {Lecture {{Notes}}
  in {{Physics}}}, Vol.\ \bibinfo {volume} {522}\ (\bibinfo  {publisher}
  {{Springer Berlin Heidelberg}},\ \bibinfo {year} {1999})\BibitemShut
  {NoStop}%
\bibitem [{\citenamefont {Boehnlein}\ \emph {et~al.}(2022)\citenamefont
  {Boehnlein}, \citenamefont {Diefenthaler}, \citenamefont {Sato},
  \citenamefont {Schram}, \citenamefont {Ziegler}, \citenamefont {Fanelli},
  \citenamefont {{Hjorth-Jensen}}, \citenamefont {Horn}, \citenamefont
  {Kuchera}, \citenamefont {Lee}, \citenamefont {Nazarewicz}, \citenamefont
  {Ostroumov}, \citenamefont {Orginos}, \citenamefont {Poon}, \citenamefont
  {Wang}, \citenamefont {Scheinker}, \citenamefont {Smith},\ and\ \citenamefont
  {Pang}}]{BoDi22}%
  \BibitemOpen
  \bibfield  {author} {\bibinfo {author} {\bibfnamefont {A.}~\bibnamefont
  {Boehnlein}}, \bibinfo {author} {\bibfnamefont {M.}~\bibnamefont
  {Diefenthaler}}, \bibinfo {author} {\bibfnamefont {N.}~\bibnamefont {Sato}},
  \bibinfo {author} {\bibfnamefont {M.}~\bibnamefont {Schram}}, \bibinfo
  {author} {\bibfnamefont {V.}~\bibnamefont {Ziegler}}, \bibinfo {author}
  {\bibfnamefont {C.}~\bibnamefont {Fanelli}}, \bibinfo {author} {\bibfnamefont
  {M.}~\bibnamefont {{Hjorth-Jensen}}}, \bibinfo {author} {\bibfnamefont
  {T.}~\bibnamefont {Horn}}, \bibinfo {author} {\bibfnamefont {M.~P.}\
  \bibnamefont {Kuchera}}, \bibinfo {author} {\bibfnamefont {D.}~\bibnamefont
  {Lee}}, \bibinfo {author} {\bibfnamefont {W.}~\bibnamefont {Nazarewicz}},
  \bibinfo {author} {\bibfnamefont {P.}~\bibnamefont {Ostroumov}}, \bibinfo
  {author} {\bibfnamefont {K.}~\bibnamefont {Orginos}}, \bibinfo {author}
  {\bibfnamefont {A.}~\bibnamefont {Poon}}, \bibinfo {author} {\bibfnamefont
  {X.-N.}\ \bibnamefont {Wang}}, \bibinfo {author} {\bibfnamefont
  {A.}~\bibnamefont {Scheinker}}, \bibinfo {author} {\bibfnamefont {M.~S.}\
  \bibnamefont {Smith}},\ and\ \bibinfo {author} {\bibfnamefont {L.-G.}\
  \bibnamefont {Pang}},\ }\bibfield  {title} {\bibinfo {title} {{Colloquium} :
  {{Machine}} learning in nuclear physics},\ }\href
  {https://doi.org/10.1103/RevModPhys.94.031003} {\bibfield  {journal}
  {\bibinfo  {journal} {Rev. Mod. Phys.}\ }\textbf {\bibinfo {volume} {94}},\
  \bibinfo {pages} {031003} (\bibinfo {year} {2022})}\BibitemShut {NoStop}%
\bibitem [{\citenamefont {Keeble}\ and\ \citenamefont {Rios}(2020)}]{KeRi20}%
  \BibitemOpen
  \bibfield  {author} {\bibinfo {author} {\bibfnamefont {J.}~\bibnamefont
  {Keeble}}\ and\ \bibinfo {author} {\bibfnamefont {A.}~\bibnamefont {Rios}},\
  }\bibfield  {title} {\bibinfo {title} {Machine learning the deuteron},\
  }\href {https://doi.org/10.1016/j.physletb.2020.135743} {\bibfield  {journal}
  {\bibinfo  {journal} {Phys. Lett. B}\ }\textbf {\bibinfo {volume} {809}},\
  \bibinfo {pages} {135743} (\bibinfo {year} {2020})}\BibitemShut {NoStop}%
\bibitem [{\citenamefont {Adams}\ \emph {et~al.}(2021)\citenamefont {Adams},
  \citenamefont {Carleo}, \citenamefont {Lovato},\ and\ \citenamefont
  {Rocco}}]{AdCa21}%
  \BibitemOpen
  \bibfield  {author} {\bibinfo {author} {\bibfnamefont {C.}~\bibnamefont
  {Adams}}, \bibinfo {author} {\bibfnamefont {G.}~\bibnamefont {Carleo}},
  \bibinfo {author} {\bibfnamefont {A.}~\bibnamefont {Lovato}},\ and\ \bibinfo
  {author} {\bibfnamefont {N.}~\bibnamefont {Rocco}},\ }\bibfield  {title}
  {\bibinfo {title} {Variational {{Monte Carlo Calculations}} of {{A}} {$\leq$}
  4 {{Nuclei}} with an {{Artificial Neural-Network Correlator Ansatz}}},\
  }\href {https://doi.org/10.1103/PhysRevLett.127.022502} {\bibfield  {journal}
  {\bibinfo  {journal} {Phys. Rev. Lett.}\ }\textbf {\bibinfo {volume} {127}},\
  \bibinfo {pages} {022502} (\bibinfo {year} {2021})}\BibitemShut {NoStop}%
\bibitem [{\citenamefont {Gnech}\ \emph {et~al.}(2022)\citenamefont {Gnech},
  \citenamefont {Adams}, \citenamefont {Brawand}, \citenamefont {Carleo},
  \citenamefont {Lovato},\ and\ \citenamefont {Rocco}}]{GnAd22}%
  \BibitemOpen
  \bibfield  {author} {\bibinfo {author} {\bibfnamefont {A.}~\bibnamefont
  {Gnech}}, \bibinfo {author} {\bibfnamefont {C.}~\bibnamefont {Adams}},
  \bibinfo {author} {\bibfnamefont {N.}~\bibnamefont {Brawand}}, \bibinfo
  {author} {\bibfnamefont {G.}~\bibnamefont {Carleo}}, \bibinfo {author}
  {\bibfnamefont {A.}~\bibnamefont {Lovato}},\ and\ \bibinfo {author}
  {\bibfnamefont {N.}~\bibnamefont {Rocco}},\ }\bibfield  {title} {\bibinfo
  {title} {Nuclei with {{Up}} to ${{A}} = 6$ {{Nucleons}} with {{Artificial
  Neural Network Wave Functions}}},\ }\href
  {https://doi.org/10.1007/s00601-021-01706-0} {\bibfield  {journal} {\bibinfo
  {journal} {Few-Body Syst.}\ }\textbf {\bibinfo {volume} {63}},\ \bibinfo
  {pages} {7} (\bibinfo {year} {2022})}\BibitemShut {NoStop}%
\bibitem [{\citenamefont {Lovato}\ \emph {et~al.}(2022)\citenamefont {Lovato},
  \citenamefont {Adams}, \citenamefont {Carleo},\ and\ \citenamefont
  {Rocco}}]{LoAd22}%
  \BibitemOpen
  \bibfield  {author} {\bibinfo {author} {\bibfnamefont {A.}~\bibnamefont
  {Lovato}}, \bibinfo {author} {\bibfnamefont {C.}~\bibnamefont {Adams}},
  \bibinfo {author} {\bibfnamefont {G.}~\bibnamefont {Carleo}},\ and\ \bibinfo
  {author} {\bibfnamefont {N.}~\bibnamefont {Rocco}},\ }\bibfield  {title}
  {\bibinfo {title} {Hidden-nucleons neural-network quantum states for the
  nuclear many-body problem},\ }\href
  {https://doi.org/10.1103/PhysRevResearch.4.043178} {\bibfield  {journal}
  {\bibinfo  {journal} {Phys. Rev. Research}\ }\textbf {\bibinfo {volume}
  {4}},\ \bibinfo {pages} {043178} (\bibinfo {year} {2022})}\BibitemShut
  {NoStop}%
\bibitem [{\citenamefont {Athanassopoulos}\ \emph {et~al.}(2004)\citenamefont
  {Athanassopoulos}, \citenamefont {Mavrommatis}, \citenamefont {Gernoth},\
  and\ \citenamefont {Clark}}]{AtMa04}%
  \BibitemOpen
  \bibfield  {author} {\bibinfo {author} {\bibfnamefont {S.}~\bibnamefont
  {Athanassopoulos}}, \bibinfo {author} {\bibfnamefont {E.}~\bibnamefont
  {Mavrommatis}}, \bibinfo {author} {\bibfnamefont {K.}~\bibnamefont
  {Gernoth}},\ and\ \bibinfo {author} {\bibfnamefont {J.}~\bibnamefont
  {Clark}},\ }\bibfield  {title} {\bibinfo {title} {Nuclear mass systematics
  using neural networks},\ }\href
  {https://doi.org/10.1016/j.nuclphysa.2004.08.006} {\bibfield  {journal}
  {\bibinfo  {journal} {Nucl. Phys. A}\ }\textbf {\bibinfo {volume} {743}},\
  \bibinfo {pages} {222} (\bibinfo {year} {2004})}\BibitemShut {NoStop}%
\bibitem [{\citenamefont {Akkoyun}\ \emph {et~al.}(2013)\citenamefont
  {Akkoyun}, \citenamefont {Bayram}, \citenamefont {Kara},\ and\ \citenamefont
  {Sinan}}]{AkBa13}%
  \BibitemOpen
  \bibfield  {author} {\bibinfo {author} {\bibfnamefont {S.}~\bibnamefont
  {Akkoyun}}, \bibinfo {author} {\bibfnamefont {T.}~\bibnamefont {Bayram}},
  \bibinfo {author} {\bibfnamefont {S.~O.}\ \bibnamefont {Kara}},\ and\
  \bibinfo {author} {\bibfnamefont {A.}~\bibnamefont {Sinan}},\ }\bibfield
  {title} {\bibinfo {title} {An artificial neural network application on
  nuclear charge radii},\ }\href
  {https://doi.org/10.1088/0954-3899/40/5/055106} {\bibfield  {journal}
  {\bibinfo  {journal} {J. Phys. G: Nucl. Part. Phys.}\ }\textbf {\bibinfo
  {volume} {40}},\ \bibinfo {pages} {055106} (\bibinfo {year}
  {2013})}\BibitemShut {NoStop}%
\bibitem [{\citenamefont {Negoita}\ \emph {et~al.}(2019)\citenamefont
  {Negoita}, \citenamefont {Vary}, \citenamefont {Luecke}, \citenamefont
  {Maris}, \citenamefont {Shirokov}, \citenamefont {Shin}, \citenamefont {Kim},
  \citenamefont {Ng}, \citenamefont {Yang}, \citenamefont {Lockner},\ and\
  \citenamefont {Prabhu}}]{NeVa19}%
  \BibitemOpen
  \bibfield  {author} {\bibinfo {author} {\bibfnamefont {G.~A.}\ \bibnamefont
  {Negoita}}, \bibinfo {author} {\bibfnamefont {J.~P.}\ \bibnamefont {Vary}},
  \bibinfo {author} {\bibfnamefont {G.~R.}\ \bibnamefont {Luecke}}, \bibinfo
  {author} {\bibfnamefont {P.}~\bibnamefont {Maris}}, \bibinfo {author}
  {\bibfnamefont {A.~M.}\ \bibnamefont {Shirokov}}, \bibinfo {author}
  {\bibfnamefont {I.~J.}\ \bibnamefont {Shin}}, \bibinfo {author}
  {\bibfnamefont {Y.}~\bibnamefont {Kim}}, \bibinfo {author} {\bibfnamefont
  {E.~G.}\ \bibnamefont {Ng}}, \bibinfo {author} {\bibfnamefont
  {C.}~\bibnamefont {Yang}}, \bibinfo {author} {\bibfnamefont {M.}~\bibnamefont
  {Lockner}},\ and\ \bibinfo {author} {\bibfnamefont {G.~M.}\ \bibnamefont
  {Prabhu}},\ }\bibfield  {title} {\bibinfo {title} {Deep learning:
  {{Extrapolation}} tool for {\emph{ab initio}} nuclear theory},\ }\href
  {https://doi.org/10.1103/PhysRevC.99.054308} {\bibfield  {journal} {\bibinfo
  {journal} {Phys. Rev. C}\ }\textbf {\bibinfo {volume} {99}},\ \bibinfo
  {pages} {054308} (\bibinfo {year} {2019})}\BibitemShut {NoStop}%
\bibitem [{\citenamefont {Jiang}\ \emph {et~al.}(2019)\citenamefont {Jiang},
  \citenamefont {Hagen},\ and\ \citenamefont {Papenbrock}}]{JiHa19}%
  \BibitemOpen
  \bibfield  {author} {\bibinfo {author} {\bibfnamefont {W.~G.}\ \bibnamefont
  {Jiang}}, \bibinfo {author} {\bibfnamefont {G.}~\bibnamefont {Hagen}},\ and\
  \bibinfo {author} {\bibfnamefont {T.}~\bibnamefont {Papenbrock}},\ }\bibfield
   {title} {\bibinfo {title} {Extrapolation of nuclear structure observables
  with artificial neural networks},\ }\href
  {https://doi.org/10.1103/PhysRevC.100.054326} {\bibfield  {journal} {\bibinfo
   {journal} {Phys. Rev. C}\ }\textbf {\bibinfo {volume} {100}},\ \bibinfo
  {pages} {054326} (\bibinfo {year} {2019})}\BibitemShut {NoStop}%
\bibitem [{\citenamefont {Kn{\"o}ll}\ \emph {et~al.}(2023)\citenamefont
  {Kn{\"o}ll}, \citenamefont {Wolfgruber}, \citenamefont {Agel}, \citenamefont
  {Wenz},\ and\ \citenamefont {Roth}}]{KnoWo23}%
  \BibitemOpen
  \bibfield  {author} {\bibinfo {author} {\bibfnamefont {M.}~\bibnamefont
  {Kn{\"o}ll}}, \bibinfo {author} {\bibfnamefont {T.}~\bibnamefont
  {Wolfgruber}}, \bibinfo {author} {\bibfnamefont {M.~L.}\ \bibnamefont
  {Agel}}, \bibinfo {author} {\bibfnamefont {C.}~\bibnamefont {Wenz}},\ and\
  \bibinfo {author} {\bibfnamefont {R.}~\bibnamefont {Roth}},\ }\bibfield
  {title} {\bibinfo {title} {Machine learning for the prediction of converged
  energies from ab initio nuclear structure calculations},\ }\href
  {https://doi.org/10.1016/j.physletb.2023.137781} {\bibfield  {journal}
  {\bibinfo  {journal} {Phys. Lett. B}\ }\textbf {\bibinfo {volume} {839}},\
  \bibinfo {pages} {137781} (\bibinfo {year} {2023})}\BibitemShut {NoStop}%
\bibitem [{\citenamefont {Maris}\ \emph {et~al.}(2009)\citenamefont {Maris},
  \citenamefont {Vary},\ and\ \citenamefont {Shirokov}}]{MaVa09}%
  \BibitemOpen
  \bibfield  {author} {\bibinfo {author} {\bibfnamefont {P.}~\bibnamefont
  {Maris}}, \bibinfo {author} {\bibfnamefont {J.~P.}\ \bibnamefont {Vary}},\
  and\ \bibinfo {author} {\bibfnamefont {A.~M.}\ \bibnamefont {Shirokov}},\
  }\bibfield  {title} {\bibinfo {title} {{\emph{Ab Initio}} no-core full
  configuration calculations of light nuclei},\ }\href
  {https://doi.org/10.1103/PhysRevC.79.014308} {\bibfield  {journal} {\bibinfo
  {journal} {Phys. Rev. C}\ }\textbf {\bibinfo {volume} {79}},\ \bibinfo
  {pages} {014308} (\bibinfo {year} {2009})}\BibitemShut {NoStop}%
\bibitem [{\citenamefont {Bogner}\ \emph {et~al.}(2008)\citenamefont {Bogner},
  \citenamefont {Furnstahl}, \citenamefont {Maris}, \citenamefont {Perry},
  \citenamefont {Schwenk},\ and\ \citenamefont {Vary}}]{BoFu08}%
  \BibitemOpen
  \bibfield  {author} {\bibinfo {author} {\bibfnamefont {S.}~\bibnamefont
  {Bogner}}, \bibinfo {author} {\bibfnamefont {R.}~\bibnamefont {Furnstahl}},
  \bibinfo {author} {\bibfnamefont {P.}~\bibnamefont {Maris}}, \bibinfo
  {author} {\bibfnamefont {R.}~\bibnamefont {Perry}}, \bibinfo {author}
  {\bibfnamefont {A.}~\bibnamefont {Schwenk}},\ and\ \bibinfo {author}
  {\bibfnamefont {J.}~\bibnamefont {Vary}},\ }\bibfield  {title} {\bibinfo
  {title} {Convergence in the no-core shell model with low-momentum two-nucleon
  interactions},\ }\href {https://doi.org/10.1016/j.nuclphysa.2007.12.008}
  {\bibfield  {journal} {\bibinfo  {journal} {Nucl. Phys. A}\ }\textbf
  {\bibinfo {volume} {801}},\ \bibinfo {pages} {21} (\bibinfo {year}
  {2008})}\BibitemShut {NoStop}%
\bibitem [{\citenamefont {Roth}\ \emph {et~al.}(2011)\citenamefont {Roth},
  \citenamefont {Langhammer}, \citenamefont {Calci}, \citenamefont {Binder},\
  and\ \citenamefont {Navr{\'a}til}}]{RoLa11}%
  \BibitemOpen
  \bibfield  {author} {\bibinfo {author} {\bibfnamefont {R.}~\bibnamefont
  {Roth}}, \bibinfo {author} {\bibfnamefont {J.}~\bibnamefont {Langhammer}},
  \bibinfo {author} {\bibfnamefont {A.}~\bibnamefont {Calci}}, \bibinfo
  {author} {\bibfnamefont {S.}~\bibnamefont {Binder}},\ and\ \bibinfo {author}
  {\bibfnamefont {P.}~\bibnamefont {Navr{\'a}til}},\ }\bibfield  {title}
  {\bibinfo {title} {Similarity-{{Transformed Chiral NN}}+3{{N Interactions}}
  for the {{{\emph{Ab Initio}}}} {{Description}} of $^{12}\text{C}$ and
  $^{16}\text{O}$},\ }\href {https://doi.org/10.1103/PhysRevLett.107.072501}
  {\bibfield  {journal} {\bibinfo  {journal} {Phys. Rev. Lett.}\ }\textbf
  {\bibinfo {volume} {107}},\ \bibinfo {pages} {072501} (\bibinfo {year}
  {2011})}\BibitemShut {NoStop}%
\bibitem [{\citenamefont {Roth}\ \emph {et~al.}(2014)\citenamefont {Roth},
  \citenamefont {Calci}, \citenamefont {Langhammer},\ and\ \citenamefont
  {Binder}}]{RoCa14}%
  \BibitemOpen
  \bibfield  {author} {\bibinfo {author} {\bibfnamefont {R.}~\bibnamefont
  {Roth}}, \bibinfo {author} {\bibfnamefont {A.}~\bibnamefont {Calci}},
  \bibinfo {author} {\bibfnamefont {J.}~\bibnamefont {Langhammer}},\ and\
  \bibinfo {author} {\bibfnamefont {S.}~\bibnamefont {Binder}},\ }\bibfield
  {title} {\bibinfo {title} {Evolved chiral {{NN}}+3{{N Hamiltonians}} for
  {\emph{ab initio}} nuclear structure calculations},\ }\href
  {https://doi.org/10.1103/PhysRevC.90.024325} {\bibfield  {journal} {\bibinfo
  {journal} {Phys. Rev. C}\ }\textbf {\bibinfo {volume} {90}},\ \bibinfo
  {pages} {024325} (\bibinfo {year} {2014})}\BibitemShut {NoStop}%
\bibitem [{\citenamefont {Jurgenson}\ \emph {et~al.}(2013)\citenamefont
  {Jurgenson}, \citenamefont {Maris}, \citenamefont {Furnstahl}, \citenamefont
  {Navr{\'a}til}, \citenamefont {Ormand},\ and\ \citenamefont {Vary}}]{JuMa13}%
  \BibitemOpen
  \bibfield  {author} {\bibinfo {author} {\bibfnamefont {E.~D.}\ \bibnamefont
  {Jurgenson}}, \bibinfo {author} {\bibfnamefont {P.}~\bibnamefont {Maris}},
  \bibinfo {author} {\bibfnamefont {R.~J.}\ \bibnamefont {Furnstahl}}, \bibinfo
  {author} {\bibfnamefont {P.}~\bibnamefont {Navr{\'a}til}}, \bibinfo {author}
  {\bibfnamefont {W.~E.}\ \bibnamefont {Ormand}},\ and\ \bibinfo {author}
  {\bibfnamefont {J.~P.}\ \bibnamefont {Vary}},\ }\bibfield  {title} {\bibinfo
  {title} {Structure of p-shell nuclei using three-nucleon interactions evolved
  with the similarity renormalization group},\ }\href
  {https://doi.org/10.1103/PhysRevC.87.054312} {\bibfield  {journal} {\bibinfo
  {journal} {Phys. Rev. C}\ }\textbf {\bibinfo {volume} {87}},\ \bibinfo
  {pages} {054312} (\bibinfo {year} {2013})}\BibitemShut {NoStop}%
\bibitem [{\citenamefont {Maris}\ \emph {et~al.}(2021)\citenamefont {Maris},
  \citenamefont {Epelbaum}, \citenamefont {Furnstahl}, \citenamefont {Golak},
  \citenamefont {Hebeler}, \citenamefont {H{\"u}ther}, \citenamefont {Kamada},
  \citenamefont {Krebs}, \citenamefont {Mei{\ss}ner}, \citenamefont {Melendez},
  \citenamefont {Nogga}, \citenamefont {Reinert}, \citenamefont {Roth},
  \citenamefont {Skibi{\'n}ski}, \citenamefont {Soloviov}, \citenamefont
  {Topolnicki}, \citenamefont {Vary}, \citenamefont {Volkotrub}, \citenamefont
  {Wita{\l}a},\ and\ \citenamefont {Wolfgruber}}]{LENPIC21}%
  \BibitemOpen
  \bibfield  {author} {\bibinfo {author} {\bibfnamefont {P.}~\bibnamefont
  {Maris}}, \bibinfo {author} {\bibfnamefont {E.}~\bibnamefont {Epelbaum}},
  \bibinfo {author} {\bibfnamefont {R.~J.}\ \bibnamefont {Furnstahl}}, \bibinfo
  {author} {\bibfnamefont {J.}~\bibnamefont {Golak}}, \bibinfo {author}
  {\bibfnamefont {K.}~\bibnamefont {Hebeler}}, \bibinfo {author} {\bibfnamefont
  {T.}~\bibnamefont {H{\"u}ther}}, \bibinfo {author} {\bibfnamefont
  {H.}~\bibnamefont {Kamada}}, \bibinfo {author} {\bibfnamefont
  {H.}~\bibnamefont {Krebs}}, \bibinfo {author} {\bibfnamefont {U.-G.}\
  \bibnamefont {Mei{\ss}ner}}, \bibinfo {author} {\bibfnamefont {J.~A.}\
  \bibnamefont {Melendez}}, \bibinfo {author} {\bibfnamefont {A.}~\bibnamefont
  {Nogga}}, \bibinfo {author} {\bibfnamefont {P.}~\bibnamefont {Reinert}},
  \bibinfo {author} {\bibfnamefont {R.}~\bibnamefont {Roth}}, \bibinfo {author}
  {\bibfnamefont {R.}~\bibnamefont {Skibi{\'n}ski}}, \bibinfo {author}
  {\bibfnamefont {V.}~\bibnamefont {Soloviov}}, \bibinfo {author}
  {\bibfnamefont {K.}~\bibnamefont {Topolnicki}}, \bibinfo {author}
  {\bibfnamefont {J.~P.}\ \bibnamefont {Vary}}, \bibinfo {author}
  {\bibfnamefont {{\relax Yu}.}~\bibnamefont {Volkotrub}}, \bibinfo {author}
  {\bibfnamefont {H.}~\bibnamefont {Wita{\l}a}},\ and\ \bibinfo {author}
  {\bibfnamefont {T.}~\bibnamefont {Wolfgruber}} (\bibinfo {collaboration}
  {LENPIC Collaboration}),\ }\bibfield  {title} {\bibinfo {title} {Light nuclei
  with semilocal momentum-space regularized chiral interactions up to third
  order},\ }\href {https://doi.org/10.1103/PhysRevC.103.054001} {\bibfield
  {journal} {\bibinfo  {journal} {Phys. Rev. C}\ }\textbf {\bibinfo {volume}
  {103}},\ \bibinfo {pages} {054001} (\bibinfo {year} {2021})}\BibitemShut
  {NoStop}%
\bibitem [{\citenamefont {Navr{\'a}til}\ \emph {et~al.}(2000)\citenamefont
  {Navr{\'a}til}, \citenamefont {Kamuntavi{\v c}ius},\ and\ \citenamefont
  {Barrett}}]{NaKa00}%
  \BibitemOpen
  \bibfield  {author} {\bibinfo {author} {\bibfnamefont {P.}~\bibnamefont
  {Navr{\'a}til}}, \bibinfo {author} {\bibfnamefont {G.~P.}\ \bibnamefont
  {Kamuntavi{\v c}ius}},\ and\ \bibinfo {author} {\bibfnamefont {B.~R.}\
  \bibnamefont {Barrett}},\ }\bibfield  {title} {\bibinfo {title} {Few-nucleon
  systems in a translationally invariant harmonic oscillator basis},\ }\href
  {https://doi.org/10.1103/PhysRevC.61.044001} {\bibfield  {journal} {\bibinfo
  {journal} {Phys. Rev. C}\ }\textbf {\bibinfo {volume} {61}},\ \bibinfo
  {pages} {044001} (\bibinfo {year} {2000})}\BibitemShut {NoStop}%
\bibitem [{\citenamefont {Nair}\ and\ \citenamefont {Hinton}(2010)}]{NaHi10}%
  \BibitemOpen
  \bibfield  {author} {\bibinfo {author} {\bibfnamefont {V.}~\bibnamefont
  {Nair}}\ and\ \bibinfo {author} {\bibfnamefont {G.~E.}\ \bibnamefont
  {Hinton}},\ }\bibfield  {title} {\bibinfo {title} {Rectified {{Linear Units
  Improve Restricted Boltzmann Machines}}},\ }in\ \href@noop {} {\emph
  {\bibinfo {booktitle} {Proceedings of the 27th {{International Conference}}
  on {{Machine Learning}}}}}\ (\bibinfo  {publisher} {{Omnipress}},\ \bibinfo
  {address} {Madison, WI, USA},\ \bibinfo {year} {2010})\ pp.\ \bibinfo {pages}
  {807--814}\BibitemShut {NoStop}%
\bibitem [{\citenamefont {Loshchilov}\ and\ \citenamefont
  {Hutter}(2017)}]{LoHu17}%
  \BibitemOpen
  \bibfield  {author} {\bibinfo {author} {\bibfnamefont {I.}~\bibnamefont
  {Loshchilov}}\ and\ \bibinfo {author} {\bibfnamefont {F.}~\bibnamefont
  {Hutter}},\ }\href@noop {} {\bibinfo {title} {Decoupled weight decay
  regularization}} (\bibinfo {year} {2017}),\ \Eprint
  {https://arxiv.org/abs/1711.05101} {arXiv:1711.05101} \BibitemShut {NoStop}%
\bibitem [{\citenamefont {Wendt}\ \emph {et~al.}(2015)\citenamefont {Wendt},
  \citenamefont {Forss{\'e}n}, \citenamefont {Papenbrock},\ and\ \citenamefont
  {S{\"a}{\"a}f}}]{WeFo15}%
  \BibitemOpen
  \bibfield  {author} {\bibinfo {author} {\bibfnamefont {K.~A.}\ \bibnamefont
  {Wendt}}, \bibinfo {author} {\bibfnamefont {C.}~\bibnamefont {Forss{\'e}n}},
  \bibinfo {author} {\bibfnamefont {T.}~\bibnamefont {Papenbrock}},\ and\
  \bibinfo {author} {\bibfnamefont {D.}~\bibnamefont {S{\"a}{\"a}f}},\
  }\bibfield  {title} {\bibinfo {title} {Infrared length scale and
  extrapolations for the no-core shell model},\ }\href
  {https://doi.org/10.1103/PhysRevC.91.061301} {\bibfield  {journal} {\bibinfo
  {journal} {Phys. Rev. C}\ }\textbf {\bibinfo {volume} {91}},\ \bibinfo
  {pages} {061301} (\bibinfo {year} {2015})}\BibitemShut {NoStop}%
\bibitem [{\citenamefont {Forss{\'e}n}\ \emph {et~al.}(2018)\citenamefont
  {Forss{\'e}n}, \citenamefont {Carlsson}, \citenamefont {Johansson},
  \citenamefont {S{\"a}{\"a}f}, \citenamefont {Bansal}, \citenamefont {Hagen},\
  and\ \citenamefont {Papenbrock}}]{FoCa18}%
  \BibitemOpen
  \bibfield  {author} {\bibinfo {author} {\bibfnamefont {C.}~\bibnamefont
  {Forss{\'e}n}}, \bibinfo {author} {\bibfnamefont {B.~D.}\ \bibnamefont
  {Carlsson}}, \bibinfo {author} {\bibfnamefont {H.~T.}\ \bibnamefont
  {Johansson}}, \bibinfo {author} {\bibfnamefont {D.}~\bibnamefont
  {S{\"a}{\"a}f}}, \bibinfo {author} {\bibfnamefont {A.}~\bibnamefont
  {Bansal}}, \bibinfo {author} {\bibfnamefont {G.}~\bibnamefont {Hagen}},\ and\
  \bibinfo {author} {\bibfnamefont {T.}~\bibnamefont {Papenbrock}},\ }\bibfield
   {title} {\bibinfo {title} {Large-scale exact diagonalizations reveal
  low-momentum scales of nuclei},\ }\href
  {https://doi.org/10.1103/PhysRevC.97.034328} {\bibfield  {journal} {\bibinfo
  {journal} {Phys. Rev. C}\ }\textbf {\bibinfo {volume} {97}},\ \bibinfo
  {pages} {034328} (\bibinfo {year} {2018})}\BibitemShut {NoStop}%
\bibitem [{\citenamefont {Entem}\ \emph {et~al.}(2017)\citenamefont {Entem},
  \citenamefont {Machleidt},\ and\ \citenamefont {Nosyk}}]{EnMa17}%
  \BibitemOpen
  \bibfield  {author} {\bibinfo {author} {\bibfnamefont {D.~R.}\ \bibnamefont
  {Entem}}, \bibinfo {author} {\bibfnamefont {R.}~\bibnamefont {Machleidt}},\
  and\ \bibinfo {author} {\bibfnamefont {Y.}~\bibnamefont {Nosyk}},\ }\bibfield
   {title} {\bibinfo {title} {High-quality two-nucleon potentials up to fifth
  order of the chiral expansion},\ }\href
  {https://doi.org/10.1103/PhysRevC.96.024004} {\bibfield  {journal} {\bibinfo
  {journal} {Phys. Rev. C}\ }\textbf {\bibinfo {volume} {96}},\ \bibinfo
  {pages} {024004} (\bibinfo {year} {2017})}\BibitemShut {NoStop}%
\bibitem [{\citenamefont {H{\"u}ther}\ \emph {et~al.}(2020)\citenamefont
  {H{\"u}ther}, \citenamefont {Vobig}, \citenamefont {Hebeler}, \citenamefont
  {Machleidt},\ and\ \citenamefont {Roth}}]{HuVo20}%
  \BibitemOpen
  \bibfield  {author} {\bibinfo {author} {\bibfnamefont {T.}~\bibnamefont
  {H{\"u}ther}}, \bibinfo {author} {\bibfnamefont {K.}~\bibnamefont {Vobig}},
  \bibinfo {author} {\bibfnamefont {K.}~\bibnamefont {Hebeler}}, \bibinfo
  {author} {\bibfnamefont {R.}~\bibnamefont {Machleidt}},\ and\ \bibinfo
  {author} {\bibfnamefont {R.}~\bibnamefont {Roth}},\ }\bibfield  {title}
  {\bibinfo {title} {Family of chiral two- plus three-nucleon interactions for
  accurate nuclear structure studies},\ }\href
  {https://doi.org/10.1016/j.physletb.2020.135651} {\bibfield  {journal}
  {\bibinfo  {journal} {Phys. Lett. B}\ }\textbf {\bibinfo {volume} {808}},\
  \bibinfo {pages} {135651} (\bibinfo {year} {2020})}\BibitemShut {NoStop}%
\bibitem [{\citenamefont {Reinert}\ \emph {et~al.}(2018)\citenamefont
  {Reinert}, \citenamefont {Krebs},\ and\ \citenamefont {Epelbaum}}]{ReKr18}%
  \BibitemOpen
  \bibfield  {author} {\bibinfo {author} {\bibfnamefont {P.}~\bibnamefont
  {Reinert}}, \bibinfo {author} {\bibfnamefont {H.}~\bibnamefont {Krebs}},\
  and\ \bibinfo {author} {\bibfnamefont {E.}~\bibnamefont {Epelbaum}},\
  }\bibfield  {title} {\bibinfo {title} {Semilocal momentum-space regularized
  chiral two-nucleon potentials up to fifth order},\ }\href
  {https://doi.org/10.1140/epja/i2018-12516-4} {\bibfield  {journal} {\bibinfo
  {journal} {Eur. Phys. J. A}\ }\textbf {\bibinfo {volume} {54}},\ \bibinfo
  {pages} {86} (\bibinfo {year} {2018})}\BibitemShut {NoStop}%
\bibitem [{\citenamefont {Maris}\ \emph {et~al.}(2023)\citenamefont {Maris},
  \citenamefont {Le}, \citenamefont {Nogga}, \citenamefont {Roth},\ and\
  \citenamefont {Vary}}]{MaLe23}%
  \BibitemOpen
  \bibfield  {author} {\bibinfo {author} {\bibfnamefont {P.}~\bibnamefont
  {Maris}}, \bibinfo {author} {\bibfnamefont {H.}~\bibnamefont {Le}}, \bibinfo
  {author} {\bibfnamefont {A.}~\bibnamefont {Nogga}}, \bibinfo {author}
  {\bibfnamefont {R.}~\bibnamefont {Roth}},\ and\ \bibinfo {author}
  {\bibfnamefont {J.~P.}\ \bibnamefont {Vary}},\ }\bibfield  {title} {\bibinfo
  {title} {Uncertainties in ab initio nuclear structure calculations with
  chiral interactions},\ }\href {https://doi.org/10.3389/fphy.2023.1098262}
  {\bibfield  {journal} {\bibinfo  {journal} {Front. Phys.}\ }\textbf {\bibinfo
  {volume} {11}},\ \bibinfo {pages} {1098262} (\bibinfo {year}
  {2023})}\BibitemShut {NoStop}%
\bibitem [{\citenamefont {Maris}()}]{MaPriv}%
  \BibitemOpen
  \bibfield  {author} {\bibinfo {author} {\bibfnamefont {P.}~\bibnamefont
  {Maris}},\ }\href@noop {} {}\bibinfo {howpublished} {private
  communication}\BibitemShut {NoStop}%
\end{thebibliography}%

\end{document}